\documentclass[11pt]{article}

\usepackage{amsthm}
\usepackage{amsmath,amssymb}          
 
\usepackage{xspace} 
\usepackage{color}
\usepackage{epsfig}    
\graphicspath{{.}{./figures/}}  

\usepackage{macros/tikzfig}
\usepackage{keycommand}

\newkeycommand{\boxmap}[style=small box][1]{\,\tikz{\node[style=\commandkey{style}] (x) {$#1$};\draw(0,-1.25)--(x)--(0,1.25);}\,}
\newkeycommand{\dboxmap}[style=dbox][1]{\,\tikz{\node[style=\commandkey{style}] (x) {$#1$};\draw[boldedge] (0,-1.25)--(x)--(0,1.25);}\,}

\newkeycommand{\wirelabel}[style=white label]{\,\tikz{\node[style=\commandkey{style}]{};}\,\xspace}

\newkeycommand{\pointmap}[style=point][1]{\,\tikz{\node[style=\commandkey{style}] (x) at (0,-0.3) {$#1$};\node [style=none] (3) at (0, 0.9) {};\node [style=none] (3) at (0, -0.9) {};\draw (x)--(0,0.7);}\,}

\newkeycommand{\point}[style=point,estyle=][1]{\,\tikz{\node[style=\commandkey{style}] (x) at (0,-0.05) {$#1$};\draw [\commandkey{estyle}] (x)--(0,1.05);}\,}

\newkeycommand{\copointmap}[style=copoint][1]{\,\tikz{\node[style=\commandkey{style}] (x) at (0,0.3) {$#1$};\draw (x)--(0,-0.7);}\,}

\newkeycommand{\dpoint}[style=point][1]{\,\tikz{\node[style=\commandkey{style},doubled] (x) at (0,-0.3) {$#1$};\draw[boldedge] (x)--(0,0.7);}\,}

\newkeycommand{\kpoint}[style=kpoint,estyle=][1]{\,\tikz{\node[style=\commandkey{style}] (x) at (0,-0.05) {$#1$};\draw [\commandkey{estyle}] (x)--(0,1.05);}\,}

\newkeycommand{\kpointgray}[style=gray kpoint,estyle=][1]{\,\tikz{\node[style=\commandkey{style}] (x) at (0,-0.05) {$#1$};\draw [\commandkey{estyle}] (x)--(0,1.05);}\,}

\newkeycommand{\typedkpoint}[style=kpoint,edgestyle=][2]{%
\,\begin{tikzpicture}
  \begin{pgfonlayer}{nodelayer}
    \node [style=none] (0) at (0, 1) {};
    \node [style=\commandkey{style}] (1) at (0, -0.75) {$#2$};
    \node [style=right label] (2) at (0.25, 0.5) {$#1$};
  \end{pgfonlayer}
  \begin{pgfonlayer}{edgelayer}
    \draw [\commandkey{edgestyle}] (1) to (0.center);
  \end{pgfonlayer}
\end{tikzpicture}\,}

\newkeycommand{\typedkpointdag}[style=kpoint adjoint,edgestyle=][2]{%
\,\begin{tikzpicture}
  \begin{pgfonlayer}{nodelayer}
    \node [style=none] (0) at (0, -1) {};
    \node [style=\commandkey{style}] (1) at (0, 0.75) {$#2$};
    \node [style=right label] (2) at (0.25, -0.5) {$#1$};
  \end{pgfonlayer}
  \begin{pgfonlayer}{edgelayer}
    \draw [\commandkey{edgestyle}] (0.center) to (1);
  \end{pgfonlayer}
\end{tikzpicture}\,}

\newkeycommand{\bistate}[style=kpoint][1]{\,\begin{tikzpicture}
  \begin{pgfonlayer}{nodelayer}
    \node [style=\commandkey{style}, minimum width=1 cm, inner sep=2pt] (0) at (0, -0.25) {$#1$};
    \node [style=none] (1) at (-0.75, 0) {};
    \node [style=none] (2) at (0.75, 0) {};
    \node [style=none] (3) at (-0.75, 0.88) {};
    \node [style=none] (4) at (0.75, 0.88) {};
  \end{pgfonlayer}
  \begin{pgfonlayer}{edgelayer}
    \draw (1.center) to (3.center);
    \draw (2.center) to (4.center);
  \end{pgfonlayer}
\end{tikzpicture}\,}

\newkeycommand{\bistatebig}[style=kpoint][1]{\,\begin{tikzpicture}
  \begin{pgfonlayer}{nodelayer}
    \node [style=\commandkey{style}, minimum width=, minimum width=1.26 cm, inner sep=2pt] (0) at (0, -0.25) {$#1$};
    \node [style=none] (1) at (-1, 0) {};
    \node [style=none] (2) at (1, 0) {};
    \node [style=none] (3) at (-1, 0.88) {};
    \node [style=none] (4) at (1, 0.88) {};
  \end{pgfonlayer}
  \begin{pgfonlayer}{edgelayer}
    \draw (1.center) to (3.center);
    \draw (2.center) to (4.center);
  \end{pgfonlayer}
\end{tikzpicture}\,}

\newkeycommand{\dbistate}[style=dkpoint][1]{\,\begin{tikzpicture}
  \begin{pgfonlayer}{nodelayer}
    \node [style=\commandkey{style}, minimum width=1 cm, inner sep=2pt] (0) at (0, -0.25) {$#1$};
    \node [style=none] (1) at (-0.75, 0) {};
    \node [style=none] (2) at (0.75, 0) {};
    \node [style=none] (3) at (-0.75, 1.0) {};
    \node [style=none] (4) at (0.75, 1.0) {};
  \end{pgfonlayer}
  \begin{pgfonlayer}{edgelayer}
    \draw [boldedge] (1.center) to (3.center);
    \draw [boldedge] (2.center) to (4.center);
  \end{pgfonlayer}
\end{tikzpicture}\,}

\newkeycommand{\bistatebraket}[style1=kpoint,style2=kpointadj][2]{\,\begin{tikzpicture}
  \begin{pgfonlayer}{nodelayer}
    \node [style=\commandkey{style1}, minimum width=1 cm, inner sep=2pt] (0) at (0, -0.75) {$#1$};
    \node [style=none] (1) at (-0.75, -0.5) {};
    \node [style=none] (2) at (0.75, -0.5) {};
    \node [style=none] (3) at (-0.75, 0.5) {};
    \node [style=none] (4) at (0.75, 0.5) {};
    \node [style=\commandkey{style2}, minimum width=1 cm, inner sep=2pt] (5) at (0, 0.75) {$#2$};
  \end{pgfonlayer}
  \begin{pgfonlayer}{edgelayer}
    \draw (1.center) to (3.center);
    \draw (2.center) to (4.center);
  \end{pgfonlayer}
\end{tikzpicture}\,}

\newkeycommand{\weight}[style=dkpoint][1]{\,\begin{tikzpicture}
  \begin{pgfonlayer}{nodelayer}
    \node [style=\commandkey{style}] (0) at (0, -0.5) {$#1$};
    \node [style=upground] (1) at (0, 0.75) {};
  \end{pgfonlayer}
  \begin{pgfonlayer}{edgelayer}
    \draw [style=boldedge] (0) to (1);
  \end{pgfonlayer}
\end{tikzpicture}\,}

\newkeycommand{\dkpoint}[style=dkpoint][1]{\,\tikz{\node[style=\commandkey{style}] (x) at (0,-0.1) {$#1$};\draw[boldedge] (x)--(0,1);}\,}
\newkeycommand{\dkpointadj}[style=dkpointadj][1]{\,\tikz{\node[style=\commandkey{style}] (x) at (0,0.1) {$#1$};\draw[boldedge] (x)--(0,-1);}\,}
\newkeycommand{\dkpointtrans}[style=dkpointtrans][1]{\,\tikz{\node[style=\commandkey{style}] (x) at (0,0.1) {$#1$};\draw[boldedge] (x)--(0,-1);}\,}
\newkeycommand{\dkpointconj}[style=dkpointconj][1]{\,\tikz{\node[style=\commandkey{style}] (x) at (0,-0.1) {$#1$};\draw[boldedge] (x)--(0,1);}\,}

\newkeycommand{\blackkpoint}[style=black kpoint][1]{\,\tikz{\node[style=\commandkey{style}] (x) at (0,-0.1) {$#1$};\draw (x)--(0,1);}\,}
\newkeycommand{\blackkpointadj}[style=black kpointadj][1]{\,\tikz{\node[style=\commandkey{style}] (x) at (0,0.1) {$#1$};\draw (x)--(0,-1);}\,}
\newkeycommand{\blackdkpoint}[style=black dkpoint][1]{\,\tikz{\node[style=\commandkey{style}] (x) at (0,-0.1) {$#1$};\draw[boldedge] (x)--(0,1);}\,}
\newkeycommand{\blackdkpointadj}[style=black dkpointadj][1]{\,\tikz{\node[style=\commandkey{style}] (x) at (0,0.1) {$#1$};\draw[boldedge] (x)--(0,-1);}\,}

\newcommand{\trace}{\,\begin{tikzpicture}
  \begin{pgfonlayer}{nodelayer}
    \node [style=none] (0) at (0, -0.60) {};
    \node [style=upground] (1) at (0, 0.40) {};
  \end{pgfonlayer}
  \begin{pgfonlayer}{edgelayer}
    \draw [style=none] (0.center) to (1);
  \end{pgfonlayer}
\end{tikzpicture}\,}

\newcommand\discard\trace

\newkeycommand{\pointketbra}[style1=copoint,style2=point,estyle=][2]{\,%
\begin{tikzpicture}
  \begin{pgfonlayer}{nodelayer}
    \node [style=none] (0) at (0, -1.75) {};
    \node [style=\commandkey{style1}] (1) at (0, -0.75) {$#1$};
    \node [style=\commandkey{style2}] (2) at (0, 0.75) {$#2$};
    \node [style=none] (3) at (0, 1.75) {};
  \end{pgfonlayer}
  \begin{pgfonlayer}{edgelayer}
    \draw [\commandkey{estyle}] (0.center) to (1);
    \draw [\commandkey{estyle}] (2) to (3.center);
  \end{pgfonlayer}
\end{tikzpicture}\,}

\newkeycommand{\twocopointketbra}[style1=copoint,style2=copoint,style3=point][3]{\,%
\begin{tikzpicture}
  \begin{pgfonlayer}{nodelayer}
    \node [style=none] (0) at (-0.75, -1.75) {};
    \node [style=none] (1) at (0.75, -1.75) {};
    \node [style=\commandkey{style1}] (2) at (-0.75, -0.75) {$#1$};
    \node [style=\commandkey{style2}] (3) at (0.75, -0.75) {$#2$};
    \node [style=\commandkey{style3}] (4) at (0, 0.75) {$#3$};
    \node [style=none] (5) at (0, 1.75) {};
  \end{pgfonlayer}
  \begin{pgfonlayer}{edgelayer}
    \draw (0.center) to (2);
    \draw (1.center) to (3);
    \draw (4) to (5.center);
  \end{pgfonlayer}
\end{tikzpicture}\,}

\newkeycommand{\twopointketbra}[style1=copoint,style2=point,style3=point][3]{\,%
\begin{tikzpicture}
  \begin{pgfonlayer}{nodelayer}
    \node [style=none] (0) at (0, -1.75) {};
    \node [style=none] (1) at (-0.75, 1.75) {};
    \node [style=\commandkey{style1}] (2) at (0, -0.75) {$#1$};
    \node [style=\commandkey{style2}] (3) at (-0.75, 0.75) {$#2$};
    \node [style=\commandkey{style3}] (4) at (0.75, 0.75) {$#3$};
    \node [style=none] (5) at (0.75, 1.75) {};
  \end{pgfonlayer}
  \begin{pgfonlayer}{edgelayer}
    \draw (0.center) to (2);
    \draw (1.center) to (3);
    \draw (4) to (5.center);
  \end{pgfonlayer}
\end{tikzpicture}\,}

\newkeycommand{\pointbraket}[style1=point,style2=copoint,estyle=][2]{\,%
\begin{tikzpicture}
  \begin{pgfonlayer}{nodelayer}
    \node [style=\commandkey{style1}] (0) at (0, -0.625) {$#1$};
    \node [style=\commandkey{style2}] (1) at (0, 0.625) {$#2$};
  \end{pgfonlayer}
  \begin{pgfonlayer}{edgelayer}
    \draw [\commandkey{estyle}] (0) to (1);
  \end{pgfonlayer}
\end{tikzpicture}\,}

\newkeycommand{\kpointketbra}[style1=kpointdag,style2=kpoint,estyle=][2]{\,%
\begin{tikzpicture}
  \begin{pgfonlayer}{nodelayer}
    \node [style=none] (0) at (0, -1.9) {};
    \node [style=\commandkey{style1}] (1) at (0, -0.95) {$#1$};
    \node [style=\commandkey{style2}] (2) at (0, 0.95) {$#2$};
    \node [style=none] (3) at (0, 1.9) {};
  \end{pgfonlayer}
  \begin{pgfonlayer}{edgelayer}
    \draw [\commandkey{estyle}] (0.center) to (1);
    \draw [\commandkey{estyle}] (2) to (3.center);
  \end{pgfonlayer}
\end{tikzpicture}\,}

\newkeycommand{\kpointmap}[style1=kpoint,style2=map][2]{\,%
\begin{tikzpicture}
  \begin{pgfonlayer}{nodelayer}
    \node [style=\commandkey{style1}] (0) at (0, -1.1) {$#1$};
    \node [style=\commandkey{style2}] (1) at (0, 0.5) {$#2$};
    \node [style=none] (2) at (0, 1.75) {};
  \end{pgfonlayer}
  \begin{pgfonlayer}{edgelayer}
    \draw (0) to (1);
    \draw (1) to (2.center);
  \end{pgfonlayer}
\end{tikzpicture}%
\,}

\newkeycommand{\kpointmaptrans}[style1=kpointconj,style2=mapadj][2]{\,%
\begin{tikzpicture}
  \begin{pgfonlayer}{nodelayer}
    \node [style=\commandkey{style1}] (0) at (0, -1.1) {$#1$};
    \node [style=\commandkey{style2}] (1) at (0, 0.5) {$#2$};
    \node [style=none] (2) at (0, 1.75) {};
  \end{pgfonlayer}
  \begin{pgfonlayer}{edgelayer}
    \draw (0) to (1);
    \draw (1) to (2.center);
  \end{pgfonlayer}
\end{tikzpicture}%
\,}

\newkeycommand{\kpointmapadj}[style1=kpointadj,style2=map][2]{\,%
\begin{tikzpicture}
  \begin{pgfonlayer}{nodelayer}
    \node [style=\commandkey{style1}] (0) at (0, 1.1) {$#1$};
    \node [style=\commandkey{style2}] (1) at (0, -0.5) {$#2$};
    \node [style=none] (2) at (0, -1.75) {};
  \end{pgfonlayer}
  \begin{pgfonlayer}{edgelayer}
    \draw (0) to (1);
    \draw (1) to (2.center);
  \end{pgfonlayer}
\end{tikzpicture}%
\,}

\newkeycommand{\dkpointmap}[style1=dkpoint,style2=dmap][2]{\,%
\begin{tikzpicture}
  \begin{pgfonlayer}{nodelayer}
    \node [style=\commandkey{style1}] (0) at (0, -1.2) {$#1$};
    \node [style=\commandkey{style2}] (1) at (0, 0.4) {$#2$};
    \node [style=none] (2) at (0, 1.7) {};
  \end{pgfonlayer}
  \begin{pgfonlayer}{edgelayer}
    \draw[style=boldedge] (0) to (1);
    \draw[style=boldedge] (1) to (2.center);
  \end{pgfonlayer}
\end{tikzpicture}%
\,}

\newkeycommand{\dkpointmapx}[style1=dkpoint,style2=dmap][2]{\,%
\begin{tikzpicture}
  \begin{pgfonlayer}{nodelayer}
    \node [style=\commandkey{style1}] (0) at (0, -1.4) {$#1$};
    \node [style=\commandkey{style2}] (1) at (0, 0.6) {$#2$};
    \node [style=none] (2) at (0, 1.9) {};
  \end{pgfonlayer}
  \begin{pgfonlayer}{edgelayer}
    \draw[style=boldedge] (0) to (1);
    \draw[style=boldedge] (1) to (2.center);
  \end{pgfonlayer}
\end{tikzpicture}%
\,}

\newkeycommand{\boxcopointmapdag}[style1=map,style2=copoint][2]{\,%
\begin{tikzpicture}
  \begin{pgfonlayer}{nodelayer}
    \node [style=none] (0) at (0, -1.75) {};
    \node [style=\commandkey{style1}] (1) at (0, -0.5) {$#1$};
    \node [style=\commandkey{style2}] (2) at (0, 1) {$#2$};
  \end{pgfonlayer}
  \begin{pgfonlayer}{edgelayer}
    \draw (0.center) to (1);
    \draw (1) to (2);
  \end{pgfonlayer}
\end{tikzpicture}%
\,}

\newkeycommand{\kpointdagmap}[style1=map,style2=kpointdag][2]{\,%
\begin{tikzpicture}
  \begin{pgfonlayer}{nodelayer}
    \node [style=none] (0) at (0, -1.75) {};
    \node [style=\commandkey{style1}] (1) at (0, -0.5) {$#1$};
    \node [style=\commandkey{style2}] (2) at (0, 1.1) {$#2$};
  \end{pgfonlayer}
  \begin{pgfonlayer}{edgelayer}
    \draw (0.center) to (1);
    \draw (1) to (2);
  \end{pgfonlayer}
\end{tikzpicture}%
\,}

\newkeycommand{\sandwichmap}[style1=point,style2=map,style3=copoint][3]{\,%
\begin{tikzpicture}
  \begin{pgfonlayer}{nodelayer}
    \node [style=\commandkey{style1}] (0) at (0, -1.50) {$#1$};
    \node [style=\commandkey{style2}] (1) at (0, 0) {$#2$};
    \node [style=\commandkey{style3}] (2) at (0, 1.50) {$#3$};
  \end{pgfonlayer}
  \begin{pgfonlayer}{edgelayer}
    \draw (0) to (1);
    \draw (1) to (2);
  \end{pgfonlayer}
\end{tikzpicture}%
}

\newkeycommand{\kpointsandwichmap}[style1=kpoint,style2=map,style3=kpointdag][3]{\,%
\begin{tikzpicture}
  \begin{pgfonlayer}{nodelayer}
    \node [style=\commandkey{style1}] (0) at (0, -1.5) {$#1$};
    \node [style=\commandkey{style2}] (1) at (0, 0) {$#2$};
    \node [style=\commandkey{style3}] (2) at (0, 1.5) {$#3$};
  \end{pgfonlayer}
  \begin{pgfonlayer}{edgelayer}
    \draw (0) to (1);
    \draw (1) to (2);
  \end{pgfonlayer}
\end{tikzpicture}%
}

\newkeycommand{\sandwichmapconj}[style1=point,style2=mapconj,style3=copoint][3]{\,%
\begin{tikzpicture}
  \begin{pgfonlayer}{nodelayer}
    \node [style=\commandkey{style1}] (0) at (0, -1.50) {$#1$};
    \node [style=\commandkey{style2}] (1) at (0, 0) {$#2$};
    \node [style=\commandkey{style3}] (2) at (0, 1.50) {$#3$};
  \end{pgfonlayer}
  \begin{pgfonlayer}{edgelayer}
    \draw (0) to (1);
    \draw (1) to (2);
  \end{pgfonlayer}
\end{tikzpicture}%
}

\newkeycommand{\sandwichtwo}[style1=point,style2=map,style3=map,style4=copoint][4]{\,%
\begin{tikzpicture}
  \begin{pgfonlayer}{nodelayer}
    \node [style=\commandkey{style2}] (0) at (0, -1) {$#2$};
    \node [style=\commandkey{style3}] (1) at (0, 1) {$#3$};
    \node [style=\commandkey{style1}] (2) at (0, -2.6) {$#1$};
    \node [style=\commandkey{style4}] (3) at (0, 2.6) {$#4$};
  \end{pgfonlayer}
  \begin{pgfonlayer}{edgelayer}
    \draw (2) to (0);
    \draw (0) to (1);
    \draw (1) to (3);
  \end{pgfonlayer}
\end{tikzpicture}\,}

\newkeycommand{\longbraket}[style1=point,style2=copoint][2]{\,%
\begin{tikzpicture}
  \begin{pgfonlayer}{nodelayer}
    \node [style=\commandkey{style1}] (0) at (0, -2.6) {$#1$};
    \node [style=\commandkey{style2}] (1) at (0, 2.6) {$#2$};
  \end{pgfonlayer}
  \begin{pgfonlayer}{edgelayer}
    \draw (0) to (1);
  \end{pgfonlayer}
\end{tikzpicture}\,}

\newkeycommand{\onb}[style=point]{\ensuremath{\{\,\tikz{\node[style=\commandkey{style}] (x) at (0,-0.3) {$j$};\draw (x)--(0,0.7);}\,\}}\xspace}

\newkeycommand{\phase}[style=white phase dot][1]{\,\begin{tikzpicture}
    \begin{pgfonlayer}{nodelayer}
        \node [style=none] (0) at (0, 1) {};
        \node [style=\commandkey{style}] (2) at (0, -0) {$#1$}; 
        \node [style=none] (3) at (0, -1) {};
    \end{pgfonlayer}
    \begin{pgfonlayer}{edgelayer}
        \draw (2) to (0.center);
        \draw (3.center) to (2);
    \end{pgfonlayer}
\end{tikzpicture}\,}

\newkeycommand{\dphase}[style=white phase ddot][1]{\,\begin{tikzpicture}
    \begin{pgfonlayer}{nodelayer}
        \node [style=none] (0) at (0, 1.25) {};
        \node [style=\commandkey{style}] (2) at (0, -0) {$#1$};
        \node [style=none] (3) at (0, -1.25) {};
    \end{pgfonlayer}
    \begin{pgfonlayer}{edgelayer}
        \draw [boldedge] (2) to (0.center);
        \draw [boldedge] (3.center) to (2);
    \end{pgfonlayer}
\end{tikzpicture}\,}

\newkeycommand{\dphasepoint}[style=white phase ddot][1]{\,\begin{tikzpicture}[yshift=-3mm]
  \begin{pgfonlayer}{nodelayer}
    \node [style=none] (0) at (0, 1) {};
    \node [style=\commandkey{style}] (1) at (0, 0) {$#1$};
  \end{pgfonlayer}
  \begin{pgfonlayer}{edgelayer}
    \draw [style=boldedge] (0.center) to (1);
  \end{pgfonlayer}
\end{tikzpicture}\,}

\newkeycommand{\dphasepointgray}[style=gray phase ddot][1]{\,\begin{tikzpicture}[yshift=-3mm]
  \begin{pgfonlayer}{nodelayer}
    \node [style=none] (0) at (0, 1) {};
    \node [style=\commandkey{style}] (1) at (0, 0) {$#1$};
  \end{pgfonlayer}
  \begin{pgfonlayer}{edgelayer}
    \draw [style=boldedge] (0.center) to (1);
  \end{pgfonlayer}
\end{tikzpicture}\,}

\newkeycommand{\dphasecopoint}[style=white phase ddot][1]{\,\begin{tikzpicture}[yshift=3mm,yscale=-1]
  \begin{pgfonlayer}{nodelayer}
    \node [style=none] (0) at (0, 1) {};
    \node [style=\commandkey{style}] (1) at (0, 0) {$#1$};
  \end{pgfonlayer}
  \begin{pgfonlayer}{edgelayer}
    \draw [style=boldedge] (0.center) to (1);
  \end{pgfonlayer}
\end{tikzpicture}\,}

\newkeycommand{\dphasepointsm}[style=white phase ddot][1]{\,\begin{tikzpicture}[yshift=-3mm]
  \begin{pgfonlayer}{nodelayer}
    \node [style=none] (0) at (0, 1) {};
    \node [style=\commandkey{style}] (1) at (0, 0) {\footnotesize\!$#1$\!};
  \end{pgfonlayer}
  \begin{pgfonlayer}{edgelayer}
    \draw [style=boldedge] (0.center) to (1);
  \end{pgfonlayer}
\end{tikzpicture}\,}

\newkeycommand{\phasepoint}[style=white phase dot][1]{\,\begin{tikzpicture}[yshift=-3mm]
  \begin{pgfonlayer}{nodelayer}
    \node [style=none] (0) at (0, 1) {};
    \node [style=\commandkey{style}] (1) at (0, 0) {$#1$};
  \end{pgfonlayer}
  \begin{pgfonlayer}{edgelayer}
    \draw (0.center) to (1);
  \end{pgfonlayer}
\end{tikzpicture}\,}

\newkeycommand{\phasecopoint}[style=white phase dot][1]{\,\begin{tikzpicture}[yshift=3mm]
  \begin{pgfonlayer}{nodelayer}
    \node [style=none] (0) at (0, -1) {};
    \node [style=\commandkey{style}] (1) at (0, 0) {$#1$};
  \end{pgfonlayer}
  \begin{pgfonlayer}{edgelayer}
    \draw (0.center) to (1);
  \end{pgfonlayer}
\end{tikzpicture}\,}

\newkeycommand{\kpointbraket}[style1=kpoint,style2=kpoint adjoint,estyle=][2]{\,%
\begin{tikzpicture}
  \begin{pgfonlayer}{nodelayer}
    \node [style=\commandkey{style1}] (0) at (0, -0.735) {$#1$};
    \node [style=\commandkey{style2}] (1) at (0, 0.735) {$#2$};
  \end{pgfonlayer}
  \begin{pgfonlayer}{edgelayer}
    \draw [\commandkey{estyle}] (0) to (1);
  \end{pgfonlayer}
\end{tikzpicture}\,}

\newkeycommand{\kkpointbraket}[style1=kpoint,style2=kpoint adjoint,estyle=][2]{\,%
\begin{tikzpicture}
  \begin{pgfonlayer}{nodelayer}
    \node [style=\commandkey{style1}] (0) at (0, -0.685) {$#1$};
    \node [style=\commandkey{style2}] (1) at (0, 0.685) {$#2$};
  \end{pgfonlayer}
  \begin{pgfonlayer}{edgelayer}
    \draw [\commandkey{estyle}] (0) to (1);
  \end{pgfonlayer}
\end{tikzpicture}\,}

\newkeycommand{\kpointbraketx}[style1=kpoint,style2=kpoint adjoint,estyle=][2]{\,%
\begin{tikzpicture}
  \begin{pgfonlayer}{nodelayer}
    \node [style=\commandkey{style1}] (0) at (0, -0.885) {$#1$};
    \node [style=\commandkey{style2}] (1) at (0, 0.885) {$#2$};
  \end{pgfonlayer}
  \begin{pgfonlayer}{edgelayer}
    \draw [\commandkey{estyle}] (0) to (1);
  \end{pgfonlayer}
\end{tikzpicture}\,}

\tikzstyle{cdiag}=[matrix of math nodes, row sep=3em, column sep=3em, text height=1.5ex, text depth=0.25ex,inner sep=0.5em]
\tikzstyle{arrow above}=[transform canvas={yshift=0.5ex}]
\tikzstyle{arrow below}=[transform canvas={yshift=-0.5ex}]

\usepackage{tikz}
\usetikzlibrary{decorations.markings}
\usetikzlibrary{shapes.geometric}

\pgfdeclarelayer{edgelayer}
\pgfdeclarelayer{nodelayer}
\pgfsetlayers{edgelayer,nodelayer,main}

\tikzstyle{none}=[inner sep=0pt]
\definecolor{hexcolor0xff0000}{rgb}{1.000,0.000,0.000}
\definecolor{hexcolor0x000000}{rgb}{0.000,0.000,0.000}
\definecolor{hexcolor0x00ff00}{rgb}{0.000,1.000,0.000}
\definecolor{hexcolor0x000000}{rgb}{0.000,0.000,0.000}
\definecolor{hexcolor0xffff00}{rgb}{1.000,1.000,0.000}
\definecolor{hexcolor0xffffff}{rgb}{1.000,1.000,1.000}

\tikzstyle{rn}=[circle,fill=hexcolor0xff0000,draw=hexcolor0x000000,line width=0.8 pt]
\tikzstyle{gn}=[circle,fill=hexcolor0x00ff00,draw=hexcolor0x000000,line width=0.8 pt]
\tikzstyle{yn}=[circle,fill=hexcolor0xffff00,draw=hexcolor0x000000,line width=0.8 pt]
\tikzstyle{wn}=[circle,fill=hexcolor0xffffff,draw=hexcolor0x000000,line width=0.8 pt]
\tikzstyle{wnthick}=[circle,fill=hexcolor0xffffff,draw=hexcolor0x000000,line width=2.500]

\tikzstyle{simple}=[-,draw=hexcolor0x000000,line width=2.000]
\tikzstyle{arrow}=[-,draw=hexcolor0x000000,postaction={decorate},decoration={markings,mark=at position .5 with {\arrow{>}}},line width=2.000]
\tikzstyle{tick}=[-,draw=hexcolor0x000000,postaction={decorate},decoration={markings,mark=at position .5 with {\draw (0,-0.1) -- (0,0.1);}},line width=2.000]
\tikzstyle{halfthickness}=[-,draw=hexcolor0x000000,line width=0.500]
\tikzstyle{thick}=[-,draw=hexcolor0x000000,line width=2.500]
\tikzstyle{thicker}=[-,draw=hexcolor0x000000,line width=4.000]

\tikzstyle{env}=[copoint,regular polygon rotate=0,minimum width=0.2cm, fill=black]

\tikzstyle{probs}=[shape=semicircle,fill=white,draw=black,shape border rotate=180,minimum width=1.2cm]

\tikzstyle{every picture}=[baseline=-0.25em,scale=0.5]
\tikzstyle{dotpic}=[] 
\tikzstyle{diredges}=[every to/.style={diredge}]
\tikzstyle{math matrix}=[matrix of math nodes,left delimiter=(,right delimiter=),inner sep=2pt,column sep=1em,row sep=0.5em,nodes={inner sep=0pt},text height=1.5ex, text depth=0.25ex]

\tikzstyle{inline text}=[text height=1.5ex, text depth=0.25ex,yshift=0.5mm]
\tikzstyle{label}=[font=\footnotesize,text height=1.5ex, text depth=0.25ex,yshift=0.5mm]
\tikzstyle{left label}=[label,anchor=east,xshift=1.5mm]
\tikzstyle{right label}=[label,anchor=west,xshift=-1.5mm]

\tikzstyle{braceedge}=[decorate,decoration={brace,amplitude=2mm,raise=-1mm}]
\tikzstyle{small braceedge}=[decorate,decoration={brace,amplitude=1mm,raise=-1mm}]

\tikzstyle{doubled}=[line width=1.6pt] 
\tikzstyle{boldedge}=[doubled,shorten <=-0.17mm,shorten >=-0.17mm]
\tikzstyle{boldedgegray}=[doubled,gray,shorten <=-0.17mm,shorten >=-0.17mm]

\tikzstyle{semidoubled}=[line width=1.4pt] 
\tikzstyle{semiboldedgegray}=[semidoubled,gray,shorten <=-0.17mm,shorten >=-0.17mm]

\tikzstyle{boldedgedashed}=[very thick,dashed,shorten <=-0.17mm,shorten >=-0.17mm]
\tikzstyle{vboldedgedashed}=[doubled,dashed,shorten <=-0.17mm,shorten >=-0.17mm]
\tikzstyle{left hook arrow}=[left hook-latex]
\tikzstyle{right hook arrow}=[right hook-latex]
\tikzstyle{sembracket}=[line width=0.5pt,shorten <=-0.07mm,shorten >=-0.07mm]

\tikzstyle{causal edge}=[->,thick,gray]
\tikzstyle{causal nondir}=[thick,gray]
\tikzstyle{timeline}=[thick,gray, dashed]

\tikzstyle{cedge}=[<->,thick,gray!70!white]

\tikzstyle{empty diagram}=[draw=gray!40!white,dashed,shape=rectangle,minimum width=1cm,minimum height=1cm]
\tikzstyle{empty diagram small}=[draw=gray!50!white,dashed,shape=rectangle,minimum width=0.6cm,minimum height=0.5cm]

\tikzstyle{dot}=[inner sep=0mm,minimum width=2mm,minimum height=2mm,draw,shape=circle]
\tikzstyle{ddot}=[inner sep=0mm, doubled, minimum width=2.5mm,minimum height=2.5mm,draw,shape=circle]

\tikzstyle{black dot}=[dot,fill=black]
\tikzstyle{white dot}=[dot,fill=white,,text depth=-0.2mm]
\tikzstyle{green dot}=[white dot] 
\tikzstyle{gray dot}=[dot,fill=gray!40!white,,text depth=-0.2mm]
\tikzstyle{red dot}=[gray dot] 

\tikzstyle{black ddot}=[ddot,fill=black]
\tikzstyle{white ddot}=[ddot,fill=white]
\tikzstyle{gray ddot}=[ddot,fill=gray!40!white]

\tikzstyle{gray edge}=[gray!40!white]

\tikzstyle{small dot}=[inner sep=0.5mm,minimum width=0pt,minimum height=0pt,draw,shape=circle]

\tikzstyle{small black dot}=[small dot,fill=black]
\tikzstyle{small white dot}=[small dot,fill=white]
\tikzstyle{small gray dot}=[small dot,fill=gray!40!white]

\tikzstyle{causal dot}=[inner sep=0.4mm,minimum width=0pt,minimum height=0pt,draw=white,shape=circle,fill=gray!40!white]

\tikzstyle{phase dimensions}=[minimum size=5mm,font=\footnotesize,rectangle,rounded corners=2.5mm,inner sep=0.2mm,outer sep=-2mm]
\tikzstyle{phase dimensions small}=[minimum size=3.0mm,font=\footnotesize,rectangle,rounded corners=1.5mm,inner sep=0.2mm,outer sep=-1.2mm]
\tikzstyle{dphase dimensions}=[minimum size=5mm,font=\footnotesize,rectangle,rounded corners=2.5mm,inner sep=0.2mm,outer sep=-2mm]

\tikzstyle{white phase dot}=[dot,fill=white,phase dimensions]
\tikzstyle{white phase dot small}=[dot,fill=white,phase dimensions small]
\tikzstyle{white phase ddot}=[ddot,fill=white,dphase dimensions]
\tikzstyle{green phase ddot}=[ddot,fill=green,dphase dimensions]

\tikzstyle{white rect ddot}=[draw=black,fill=white,doubled,minimum size=5mm,font=\footnotesize,rectangle,rounded corners=2.5mm,inner sep=0.2mm]
\tikzstyle{gray rect ddot}=[draw=black,fill=gray!40!white,doubled,minimum size=6mm,font=\footnotesize,rectangle,rounded corners=3mm]

\tikzstyle{gray phase dot}=[dot,fill=gray!40!white,phase dimensions]
\tikzstyle{gray phase dot small}=[dot,fill=gray!40!white,phase dimensions small]
\tikzstyle{gray phase ddot}=[ddot,fill=gray!40!white,dphase dimensions]
\tikzstyle{red phase ddot}=[ddot,fill=red,dphase dimensions]

\tikzstyle{grey phase dot}=[gray phase dot]
\tikzstyle{grey phase ddot}=[gray phase ddot]

\tikzstyle{small phase dimensions}=[minimum size=4mm,font=\tiny,rectangle,rounded corners=2mm,inner sep=0.2mm,outer sep=-2mm]
\tikzstyle{small dphase dimensions}=[minimum size=4mm,font=\tiny,rectangle,rounded corners=2mm,inner sep=0.2mm,outer sep=-2mm]

\tikzstyle{small gray phase dot}=[dot,fill=gray!40!white,small phase dimensions]
\tikzstyle{small gray phase ddot}=[ddot,fill=gray!40!white,small dphase dimensions]

\tikzstyle{small map}=[draw,shape=rectangle,minimum height=4mm,minimum width=4mm,fill=white]

\tikzstyle{cnot}=[fill=white,shape=circle,inner sep=-1.4pt]

\tikzstyle{asym hadamard}=[fill=white,draw,shape=NEbox,inner sep=0.6mm,font=\footnotesize,minimum height=4mm]
\tikzstyle{asym hadamard conj}=[fill=white,draw,shape=NWbox,inner sep=0.6mm,font=\footnotesize,minimum height=4mm]
\tikzstyle{asym hadamard dag}=[fill=white,draw,shape=SEbox,inner sep=0.6mm,font=\footnotesize,minimum height=4mm]

\tikzstyle{hadamard}=[fill=white,draw,inner sep=0.6mm,font=\footnotesize,minimum height=4mm,minimum width=4mm]
\tikzstyle{small hadamard}=[fill=white,draw,inner sep=0.6mm,minimum height=1.5mm,minimum width=1.5mm]
\tikzstyle{dhadamard}=[hadamard,doubled]
\tikzstyle{small dhadamard}=[small hadamard,doubled]
\tikzstyle{small dhadamard rotate}=[small hadamard,doubled,rotate=45]
\tikzstyle{antipode}=[white dot,inner sep=0.3mm,font=\footnotesize]

\tikzstyle{scalar}=[diamond,draw,inner sep=0.5pt,font=\small]
\tikzstyle{dscalar}=[diamond,doubled, draw,inner sep=0.5pt,font=\small]

\tikzstyle{small box}=[rectangle,inline text,fill=white,draw,minimum height=5mm,yshift=-0.5mm,minimum width=5mm,font=\small]
\tikzstyle{small gray box}=[small box,fill=gray!30]
\tikzstyle{medium box}=[rectangle,inline text,fill=white,draw,minimum height=5mm,yshift=-0.5mm,minimum width=10mm,font=\small]
\tikzstyle{square box}=[small box] 
\tikzstyle{medium gray box}=[small box,fill=gray!30]
\tikzstyle{semilarge box}=[rectangle,inline text,fill=white,draw,minimum height=5mm,yshift=-0.5mm,minimum width=12.5mm,font=\small]
\tikzstyle{large box}=[rectangle,inline text,fill=white,draw,minimum height=5mm,yshift=-0.5mm,minimum width=15mm,font=\small]
\tikzstyle{large gray box}=[small box,fill=gray!30]

\tikzstyle{Bayes box}=[rectangle,fill=black,draw, minimum height=3mm, minimum width=3mm]

\tikzstyle{gray square point}=[small box,fill=gray!50]

\tikzstyle{dphase box white}=[dhadamard]
\tikzstyle{dphase box gray}=[dhadamard,fill=gray!50!white]

\tikzstyle{point}=[regular polygon,regular polygon sides=3,draw,scale=0.75,inner sep=-0.5pt,minimum width=9mm,fill=white,regular polygon rotate=180]
\tikzstyle{copoint}=[regular polygon,regular polygon sides=3,draw,scale=0.75,inner sep=-0.5pt,minimum width=9mm,fill=white]
\tikzstyle{dpoint}=[point,doubled]
\tikzstyle{dcopoint}=[copoint,doubled]

\tikzstyle{wide copoint}=[fill=white,draw,shape=isosceles triangle,shape border rotate=90,isosceles triangle stretches=true,inner sep=0pt,minimum width=1.5cm,minimum height=6.12mm]
\tikzstyle{wide point}=[fill=white,draw,shape=isosceles triangle,shape border rotate=-90,isosceles triangle stretches=true,inner sep=0pt,minimum width=1.5cm,minimum height=6.12mm,yshift=-0.0mm]
\tikzstyle{wide point plus}=[fill=white,draw,shape=isosceles triangle,shape border rotate=-90,isosceles triangle stretches=true,inner sep=0pt,minimum width=1.74cm,minimum height=7mm,yshift=-0.0mm]

\tikzstyle{wide dpoint}=[fill=white,doubled,draw,shape=isosceles triangle,shape border rotate=-90,isosceles triangle stretches=true,inner sep=0pt,minimum width=1.5cm,minimum height=6.12mm,yshift=-0.0mm]
\tikzstyle{wide dcopoint}=[fill=white,doubled,draw,shape=isosceles triangle,shape border rotate=90,isosceles triangle stretches=true,inner sep=0pt,minimum width=1.5cm,minimum height=6.12mm,yshift=-0.0mm]

\tikzstyle{tinypoint}=[regular polygon,regular polygon sides=3,draw,scale=0.55,inner sep=-0.15pt,minimum width=6mm,fill=white,regular polygon rotate=180]

\tikzstyle{white point}=[point]
\tikzstyle{white dpoint}=[dpoint]
\tikzstyle{green point}=[white point] 
\tikzstyle{white copoint}=[copoint]
\tikzstyle{gray point}=[point,fill=gray!40!white]
\tikzstyle{gray dpoint}=[gray point,doubled]
\tikzstyle{red point}=[gray point] 
\tikzstyle{gray copoint}=[copoint,fill=gray!40!white]
\tikzstyle{gray dcopoint}=[gray copoint,doubled]

\tikzstyle{white point guide}=[regular polygon,regular polygon sides=3,font=\scriptsize,draw,scale=0.65,inner sep=-0.5pt,minimum width=9mm,fill=white,regular polygon rotate=180]

\tikzstyle{black point}=[point,fill=black,font=\color{white}]
\tikzstyle{black copoint}=[copoint,fill=black,font=\color{white}]

\tikzstyle{tiny gray point}=[tinypoint,fill=gray!40!white]

\tikzstyle{diredge}=[->]
\tikzstyle{ddiredge}=[<->]
\tikzstyle{rdiredge}=[<-]
\tikzstyle{thickdiredge}=[->, very thick]
\tikzstyle{pointer edge}=[->,very thick,gray]
\tikzstyle{pointer edge part}=[very thick,gray]
\tikzstyle{dashed edge}=[dashed]
\tikzstyle{thick dashed edge}=[very thick,dashed]
\tikzstyle{thick gray dashed edge}=[thick dashed edge,gray!40]
\tikzstyle{thick map edge}=[very thick,|->]

\makeatletter
\newcommand{\boxshape}[3]{%
\pgfdeclareshape{#1}{
\inheritsavedanchors[from=rectangle] 
\inheritanchorborder[from=rectangle]
\inheritanchor[from=rectangle]{center}
\inheritanchor[from=rectangle]{north}
\inheritanchor[from=rectangle]{south}
\inheritanchor[from=rectangle]{west}
\inheritanchor[from=rectangle]{east}
\backgroundpath{
\southwest \pgf@xa=\pgf@x \pgf@ya=\pgf@y
\northeast \pgf@xb=\pgf@x \pgf@yb=\pgf@y

\@tempdima=#2
\@tempdimb=#3

\pgfpathmoveto{\pgfpoint{\pgf@xa - 5pt + \@tempdima}{\pgf@ya}}
\pgfpathlineto{\pgfpoint{\pgf@xa - 5pt - \@tempdima}{\pgf@yb}}
\pgfpathlineto{\pgfpoint{\pgf@xb + 5pt + \@tempdimb}{\pgf@yb}}
\pgfpathlineto{\pgfpoint{\pgf@xb + 5pt - \@tempdimb}{\pgf@ya}}
\pgfpathlineto{\pgfpoint{\pgf@xa - 5pt + \@tempdima}{\pgf@ya}}
\pgfpathclose
}
}}

\boxshape{NEbox}{0pt}{5pt}
\boxshape{SEbox}{0pt}{-5pt}
\boxshape{NWbox}{5pt}{0pt}
\boxshape{SWbox}{-5pt}{0pt}
\boxshape{EBox}{-3pt}{3pt}
\boxshape{WBox}{3pt}{-3pt}
\makeatother

\tikzstyle{cloud}=[shape=cloud,draw,minimum width=1.5cm,minimum height=1.5cm]

\tikzstyle{map}=[draw,shape=NEbox,inner sep=2pt,minimum height=6mm,fill=white]
\tikzstyle{dashedmap}=[draw,dashed,shape=NEbox,inner sep=2pt,minimum height=6mm,fill=white]
\tikzstyle{mapdag}=[draw,shape=SEbox,inner sep=2pt,minimum height=6mm,fill=white]
\tikzstyle{mapadj}=[draw,shape=SEbox,inner sep=2pt,minimum height=6mm,fill=white]
\tikzstyle{maptrans}=[draw,shape=SWbox,inner sep=2pt,minimum height=6mm,fill=white]
\tikzstyle{mapconj}=[draw,shape=NWbox,inner sep=2pt,minimum height=6mm,fill=white]

\tikzstyle{medium map}=[draw,shape=NEbox,inner sep=2pt,minimum height=6mm,fill=white,minimum width=7mm]
\tikzstyle{medium map dag}=[draw,shape=SEbox,inner sep=2pt,minimum height=6mm,fill=white,minimum width=7mm]
\tikzstyle{medium map adj}=[draw,shape=SEbox,inner sep=2pt,minimum height=6mm,fill=white,minimum width=7mm]
\tikzstyle{medium map trans}=[draw,shape=SWbox,inner sep=2pt,minimum height=6mm,fill=white,minimum width=7mm]
\tikzstyle{medium map conj}=[draw,shape=NWbox,inner sep=2pt,minimum height=6mm,fill=white,minimum width=7mm]
\tikzstyle{semilarge map}=[draw,shape=NEbox,inner sep=2pt,minimum height=6mm,fill=white,minimum width=9.5mm]
\tikzstyle{semilarge map trans}=[draw,shape=SWbox,inner sep=2pt,minimum height=6mm,fill=white,minimum width=9.5mm]
\tikzstyle{semilarge map adj}=[draw,shape=SEbox,inner sep=2pt,minimum height=6mm,fill=white,minimum width=9.5mm]
\tikzstyle{semilarge map dag}=[draw,shape=SEbox,inner sep=2pt,minimum height=6mm,fill=white,minimum width=9.5mm]
\tikzstyle{semilarge map conj}=[draw,shape=NWbox,inner sep=2pt,minimum height=6mm,fill=white,minimum width=9.5mm]
\tikzstyle{large map}=[draw,shape=NEbox,inner sep=2pt,minimum height=6mm,fill=white,minimum width=12mm]
\tikzstyle{large map conj}=[draw,shape=NWbox,inner sep=2pt,minimum height=6mm,fill=white,minimum width=12mm]
\tikzstyle{very large map}=[draw,shape=NEbox,inner sep=2pt,minimum height=6mm,fill=white,minimum width=17mm]

\tikzstyle{medium dmap}=[draw,doubled,shape=NEbox,inner sep=2pt,minimum height=6mm,fill=white,minimum width=7mm]
\tikzstyle{medium dmap dag}=[draw,doubled,shape=SEbox,inner sep=2pt,minimum height=6mm,fill=white,minimum width=7mm]
\tikzstyle{medium dmap adj}=[draw,doubled,shape=SEbox,inner sep=2pt,minimum height=6mm,fill=white,minimum width=7mm]
\tikzstyle{medium dmap trans}=[draw,doubled,shape=SWbox,inner sep=2pt,minimum height=6mm,fill=white,minimum width=7mm]
\tikzstyle{medium dmap conj}=[draw,doubled,shape=NWbox,inner sep=2pt,minimum height=6mm,fill=white,minimum width=7mm]
\tikzstyle{semilarge dmap}=[draw,doubled,shape=NEbox,inner sep=2pt,minimum height=6mm,fill=white,minimum width=9.5mm]
\tikzstyle{semilarge dmap trans}=[draw,doubled,shape=SWbox,inner sep=2pt,minimum height=6mm,fill=white,minimum width=9.5mm]
\tikzstyle{semilarge dmap adj}=[draw,doubled,shape=SEbox,inner sep=2pt,minimum height=6mm,fill=white,minimum width=9.5mm]
\tikzstyle{semilarge dmap dag}=[draw,doubled,shape=SEbox,inner sep=2pt,minimum height=6mm,fill=white,minimum width=9.5mm]
\tikzstyle{semilarge dmap conj}=[draw,doubled,shape=NWbox,inner sep=2pt,minimum height=6mm,fill=white,minimum width=9.5mm]
\tikzstyle{large dmap}=[draw,doubled,shape=NEbox,inner sep=2pt,minimum height=6mm,fill=white,minimum width=12mm]
\tikzstyle{large dmap conj}=[draw,doubled,shape=NWbox,inner sep=2pt,minimum height=6mm,fill=white,minimum width=12mm]
\tikzstyle{large dmap trans}=[draw,doubled,shape=SWbox,inner sep=2pt,minimum height=6mm,fill=white,minimum width=12mm]
\tikzstyle{large dmap adj}=[draw,doubled,shape=SEbox,inner sep=2pt,minimum height=6mm,fill=white,minimum width=12mm]
\tikzstyle{large dmap dag}=[draw,doubled,shape=SEbox,inner sep=2pt,minimum height=6mm,fill=white,minimum width=12mm]
\tikzstyle{very large dmap}=[draw,doubled,shape=NEbox,inner sep=2pt,minimum height=6mm,fill=white,minimum width=19.5mm]

\tikzstyle{muxbox}=[draw,shape=rectangle,minimum height=3mm,minimum width=3mm,fill=white]
\tikzstyle{dmuxbox}=[muxbox,doubled]

\tikzstyle{box}=[draw,shape=rectangle,inner sep=2pt,minimum height=6mm,minimum width=6mm,fill=white]
\tikzstyle{dbox}=[draw,doubled,shape=rectangle,inner sep=2pt,minimum height=6mm,minimum width=6mm,fill=white]
\tikzstyle{dmap}=[draw,doubled,shape=NEbox,inner sep=2pt,minimum height=6mm,fill=white]
\tikzstyle{dmapdag}=[draw,doubled,shape=SEbox,inner sep=2pt,minimum height=6mm,fill=white]
\tikzstyle{dmapadj}=[draw,doubled,shape=SEbox,inner sep=2pt,minimum height=6mm,fill=white]
\tikzstyle{dmaptrans}=[draw,doubled,shape=SWbox,inner sep=2pt,minimum height=6mm,fill=white]
\tikzstyle{dmapconj}=[draw,doubled,shape=NWbox,inner sep=2pt,minimum height=6mm,fill=white]

\tikzstyle{ddmap}=[draw,doubled,dashed,shape=NEbox,inner sep=2pt,minimum height=6mm,fill=white]
\tikzstyle{ddmapdag}=[draw,doubled,dashed,shape=SEbox,inner sep=2pt,minimum height=6mm,fill=white]
\tikzstyle{ddmapadj}=[draw,doubled,dashed,shape=SEbox,inner sep=2pt,minimum height=6mm,fill=white]
\tikzstyle{ddmaptrans}=[draw,doubled,dashed,shape=SWbox,inner sep=2pt,minimum height=6mm,fill=white]
\tikzstyle{ddmapconj}=[draw,doubled,dashed,shape=NWbox,inner sep=2pt,minimum height=6mm,fill=white]

\boxshape{sNEbox}{0pt}{3pt}
\boxshape{sSEbox}{0pt}{-3pt}
\boxshape{sNWbox}{3pt}{0pt}
\boxshape{sSWbox}{-3pt}{0pt}
\tikzstyle{smap}=[draw,shape=sNEbox,fill=white]
\tikzstyle{smapdag}=[draw,shape=sSEbox,fill=white]
\tikzstyle{smapadj}=[draw,shape=sSEbox,fill=white]
\tikzstyle{smaptrans}=[draw,shape=sSWbox,fill=white]
\tikzstyle{smapconj}=[draw,shape=sNWbox,fill=white]

\tikzstyle{dsmap}=[draw,dashed,shape=sNEbox,fill=white]
\tikzstyle{dsmapdag}=[draw,dashed,shape=sSEbox,fill=white]
\tikzstyle{dsmaptrans}=[draw,dashed,shape=sSWbox,fill=white]
\tikzstyle{dsmapconj}=[draw,dashed,shape=sNWbox,fill=white]

\boxshape{mNEbox}{0pt}{10pt}
\boxshape{mSEbox}{0pt}{-10pt}
\boxshape{mNWbox}{10pt}{0pt}
\boxshape{mSWbox}{-10pt}{0pt}
\tikzstyle{mmap}=[draw,shape=mNEbox]
\tikzstyle{mmapdag}=[draw,shape=mSEbox]
\tikzstyle{mmaptrans}=[draw,shape=mSWbox]
\tikzstyle{mmapconj}=[draw,shape=mNWbox]

\tikzstyle{mmapgray}=[draw,fill=gray!40!white,shape=mNEbox]
\tikzstyle{smapgray}=[draw,fill=gray!40!white,shape=sNEbox]

\makeatletter
\pgfdeclareshape{cornerpoint}{
\inheritsavedanchors[from=rectangle] 
\inheritanchorborder[from=rectangle]
\inheritanchor[from=rectangle]{center}
\inheritanchor[from=rectangle]{north}
\inheritanchor[from=rectangle]{south}
\inheritanchor[from=rectangle]{west}
\inheritanchor[from=rectangle]{east}
\backgroundpath{
\southwest \pgf@xa=\pgf@x \pgf@ya=\pgf@y
\northeast \pgf@xb=\pgf@x \pgf@yb=\pgf@y

\pgfmathsetmacro{\pgf@shorten@left}{\pgfkeysvalueof{/tikz/shorten left}}
\pgfmathsetmacro{\pgf@shorten@right}{\pgfkeysvalueof{/tikz/shorten right}}

\pgfpathmoveto{\pgfpoint{0.5 * (\pgf@xa + \pgf@xb)}{\pgf@ya - 5pt}}
\pgfpathlineto{\pgfpoint{\pgf@xa - 8pt + \pgf@shorten@left}{\pgf@yb - 1.5 * \pgf@shorten@left}}
\pgfpathlineto{\pgfpoint{\pgf@xa - 8pt + \pgf@shorten@left}{\pgf@yb}}
\pgfpathlineto{\pgfpoint{\pgf@xb + 8pt - \pgf@shorten@right}{\pgf@yb}}
\pgfpathlineto{\pgfpoint{\pgf@xb + 8pt - \pgf@shorten@right}{\pgf@yb - 1.5 * \pgf@shorten@right}}
\pgfpathclose
}
}

\pgfdeclareshape{cornercopoint}{
\inheritsavedanchors[from=rectangle] 
\inheritanchorborder[from=rectangle]
\inheritanchor[from=rectangle]{center}
\inheritanchor[from=rectangle]{north}
\inheritanchor[from=rectangle]{south}
\inheritanchor[from=rectangle]{west}
\inheritanchor[from=rectangle]{east}
\backgroundpath{
\southwest \pgf@xa=\pgf@x \pgf@ya=\pgf@y
\northeast \pgf@xb=\pgf@x \pgf@yb=\pgf@y

\pgfmathsetmacro{\pgf@shorten@left}{\pgfkeysvalueof{/tikz/shorten left}}
\pgfmathsetmacro{\pgf@shorten@right}{\pgfkeysvalueof{/tikz/shorten right}}

\pgfpathmoveto{\pgfpoint{0.5 * (\pgf@xa + \pgf@xb)}{\pgf@yb + 5pt}}
\pgfpathlineto{\pgfpoint{\pgf@xa - 8pt + \pgf@shorten@left}{\pgf@ya + 1.5 * \pgf@shorten@left}}
\pgfpathlineto{\pgfpoint{\pgf@xa - 8pt + \pgf@shorten@left}{\pgf@ya}}
\pgfpathlineto{\pgfpoint{\pgf@xb + 8pt - \pgf@shorten@right}{\pgf@ya}}
\pgfpathlineto{\pgfpoint{\pgf@xb + 8pt - \pgf@shorten@right}{\pgf@ya + 1.5 * \pgf@shorten@right}}
\pgfpathclose
}
}

\pgfdeclareshape{langpoint}{
\inheritsavedanchors[from=rectangle] 
\inheritanchorborder[from=rectangle]
\inheritanchor[from=rectangle]{center}
\inheritanchor[from=rectangle]{north}
\inheritanchor[from=rectangle]{south}
\inheritanchor[from=rectangle]{west}
\inheritanchor[from=rectangle]{east}
\backgroundpath{
\southwest \pgf@xa=\pgf@x \pgf@ya=\pgf@y
\northeast \pgf@xb=\pgf@x \pgf@yb=\pgf@y

\pgfmathsetmacro{\pgf@shorten@left}{\pgfkeysvalueof{/tikz/shorten left}}
\pgfmathsetmacro{\pgf@shorten@right}{\pgfkeysvalueof{/tikz/shorten right}}

\pgfpathmoveto{\pgfpoint{0.5 * (\pgf@xa + \pgf@xb)}{\pgf@ya - 2pt}}
\pgfpathlineto{\pgfpoint{\pgf@xa - 8pt}{\pgf@yb - 3 * \pgf@shorten@left + 5pt}} 
\pgfpathlineto{\pgfpoint{\pgf@xa - 8pt}{\pgf@yb -1pt}}
\pgfpathlineto{\pgfpoint{\pgf@xb + 8pt}{\pgf@yb -1pt}}
\pgfpathlineto{\pgfpoint{\pgf@xb + 8pt}{\pgf@yb - 3 * \pgf@shorten@left + 5pt}}
\pgfpathclose
}
}

\pgfdeclareshape{langcopoint}{
\inheritsavedanchors[from=rectangle] 
\inheritanchorborder[from=rectangle]
\inheritanchor[from=rectangle]{center}
\inheritanchor[from=rectangle]{north}
\inheritanchor[from=rectangle]{south}
\inheritanchor[from=rectangle]{west}
\inheritanchor[from=rectangle]{east}
\backgroundpath{
\southwest \pgf@xa=\pgf@x \pgf@ya=\pgf@y
\northeast \pgf@xb=\pgf@x \pgf@yb=\pgf@y

\pgfmathsetmacro{\pgf@shorten@left}{\pgfkeysvalueof{/tikz/shorten left}}
\pgfmathsetmacro{\pgf@shorten@right}{\pgfkeysvalueof{/tikz/shorten right}}

\pgfpathmoveto{\pgfpoint{0.5 * (\pgf@xa + \pgf@xb)}{\pgf@yb +0pt}}
\pgfpathlineto{\pgfpoint{\pgf@xa - 8pt}{\pgf@ya + 3 * \pgf@shorten@left - 5pt}} 
\pgfpathlineto{\pgfpoint{\pgf@xa - 8pt}{\pgf@ya + 1pt}}
\pgfpathlineto{\pgfpoint{\pgf@xb + 8pt}{\pgf@ya + 1pt}}
\pgfpathlineto{\pgfpoint{\pgf@xb + 8pt}{\pgf@ya + 3 * \pgf@shorten@left - 5pt}}
\pgfpathclose
}
}

\pgfdeclareshape{langrect}{
\inheritsavedanchors[from=rectangle] 
\inheritanchorborder[from=rectangle]
\inheritanchor[from=rectangle]{center}
\inheritanchor[from=rectangle]{north}
\inheritanchor[from=rectangle]{south}
\inheritanchor[from=rectangle]{west}
\inheritanchor[from=rectangle]{east}
\backgroundpath{
\southwest \pgf@xa=\pgf@x \pgf@ya=\pgf@y
\northeast \pgf@xb=\pgf@x \pgf@yb=\pgf@y

\pgfmathsetmacro{\pgf@shorten@left}{\pgfkeysvalueof{/tikz/shorten left}}
\pgfmathsetmacro{\pgf@shorten@right}{\pgfkeysvalueof{/tikz/shorten right}}

\pgfpathmoveto{\pgfpoint{\pgf@xa - 8pt}{\pgf@ya + 3 * \pgf@shorten@left - 5pt}} 
\pgfpathlineto{\pgfpoint{\pgf@xa - 8pt}{\pgf@ya + 1pt}}
\pgfpathlineto{\pgfpoint{\pgf@xb + 8pt}{\pgf@ya + 1pt}}
\pgfpathlineto{\pgfpoint{\pgf@xb + 8pt}{\pgf@ya + 3 * \pgf@shorten@left - 5pt}}
\pgfpathclose
}
}

\makeatother

\pgfkeyssetvalue{/tikz/shorten left}{0pt}
\pgfkeyssetvalue{/tikz/shorten right}{0pt}

\tikzstyle{kpoint common}=[draw,fill=white,inner sep=1pt,minimum height=4mm]

\tikzstyle{langstate}=[shape=langcopoint,shorten left=5pt,kpoint common,font=\footnotesize]
\tikzstyle{langeffect}=[shape=langpoint,shorten left=5pt,kpoint common,font=\footnotesize]
\tikzstyle{langstatedash}=[shape=langcopoint,dashed, shorten left=5pt,kpoint common,font=\footnotesize]
\tikzstyle{langeffectdash}=[shape=langpoint,dashed, shorten left=5pt,kpoint common,font=\footnotesize]
\tikzstyle{langbox}=[shape=langrect,shorten left=5pt,kpoint common,font=\footnotesize] 

\tikzstyle{kpoint}=[shape=cornerpoint,shorten left=5pt,kpoint common]
\tikzstyle{kpoint adjoint}=[shape=cornercopoint,shorten left=5pt,kpoint common]

\tikzstyle{kpoint conjugate}=[shape=cornerpoint,shorten right=5pt,kpoint common]
\tikzstyle{kpoint transpose}=[shape=cornercopoint,shorten right=5pt,kpoint common]
\tikzstyle{kpoint symm}=[shape=cornerpoint,shorten left=5pt,shorten right=5pt,kpoint common]

\tikzstyle{black kpoint}=[shape=cornerpoint,shorten left=5pt,kpoint common,fill=black,font=\color{white}]
\tikzstyle{black kpoint adjoint}=[shape=cornercopoint,shorten left=5pt,kpoint common,fill=black,font=\color{white}]
\tikzstyle{black kpointadj}=[shape=cornercopoint,shorten left=5pt,kpoint common,fill=black,font=\color{white}]

\tikzstyle{black dkpoint}=[shape=cornerpoint,shorten left=5pt,kpoint common,fill=black, doubled,font=\color{white}]
\tikzstyle{black dkpoint adjoint}=[shape=cornercopoint,shorten left=5pt,kpoint common,fill=black, doubled,font=\color{white}]
\tikzstyle{black dkpointadj}=[shape=cornercopoint,shorten left=5pt,kpoint common,fill=black, doubled,font=\color{white}]

\tikzstyle{kpointdag}=[kpoint adjoint]
\tikzstyle{kpointadj}=[kpoint adjoint]
\tikzstyle{kpointconj}=[kpoint conjugate]
\tikzstyle{kpointtrans}=[kpoint transpose]

\tikzstyle{big kpoint}=[kpoint, minimum width=1.2 cm, minimum height=8mm, inner sep=4pt, text depth=3mm]

\tikzstyle{wide kpoint}=[kpoint, minimum width=1 cm, inner sep=2pt]
\tikzstyle{wide kpointdag}=[kpointdag, minimum width=1 cm, inner sep=2pt]
\tikzstyle{wide kpointconj}=[kpointconj, minimum width=1 cm, inner sep=2pt]
\tikzstyle{wide kpointtrans}=[kpointtrans, minimum width=1 cm, inner sep=2pt]

\tikzstyle{gray kpoint}=[kpoint,fill=gray!50!white]
\tikzstyle{gray kpointdag}=[kpointdag,fill=gray!50!white]
\tikzstyle{gray kpointadj}=[kpointadj,fill=gray!50!white]
\tikzstyle{gray kpointconj}=[kpointconj,fill=gray!50!white]
\tikzstyle{gray kpointtrans}=[kpointtrans,fill=gray!50!white]

\tikzstyle{gray dkpoint}=[kpoint,fill=gray!50!white,doubled]
\tikzstyle{gray dkpointdag}=[kpointdag,fill=gray!50!white,doubled]
\tikzstyle{gray dkpointadj}=[kpointadj,fill=gray!50!white,doubled]
\tikzstyle{gray dkpointconj}=[kpointconj,fill=gray!50!white,doubled]
\tikzstyle{gray dkpointtrans}=[kpointtrans,fill=gray!50!white,doubled]

\tikzstyle{white label}=[draw,fill=white,rectangle,inner sep=0.7 mm]
\tikzstyle{gray label}=[draw,fill=gray!50!white,rectangle,inner sep=0.7 mm]
\tikzstyle{black label}=[draw,fill=black,rectangle,inner sep=0.7 mm]

\tikzstyle{dkpoint}=[kpoint,doubled]
\tikzstyle{wide dkpoint}=[wide kpoint,doubled]
\tikzstyle{dkpointdag}=[kpoint adjoint,doubled]
\tikzstyle{wide dkpointdag}=[wide kpointdag,doubled]
\tikzstyle{dkcopoint}=[kpoint adjoint,doubled]
\tikzstyle{dkpointadj}=[kpoint adjoint,doubled]
\tikzstyle{dkpointconj}=[kpoint conjugate,doubled]
\tikzstyle{dkpointtrans}=[kpoint transpose,doubled]

\tikzstyle{kscalar}=[kpoint common, shape=EBox, inner xsep=-1pt, inner ysep=3pt,font=\small]
\tikzstyle{kscalarconj}=[kpoint common, shape=WBox, inner xsep=-1pt, inner ysep=3pt,font=\small]

 \tikzstyle{upground}=[circuit ee IEC,ground,rotate=90,scale=2.5]
 \tikzstyle{downground}=[circuit ee IEC,ground,rotate=-90,scale=2.5]
 \tikzstyle{bigground}=[regular polygon,regular polygon sides=3,draw=gray,scale=0.50,inner sep=-0.5pt,minimum width=10mm,fill=gray]

\tikzstyle{arrs}=[-latex,font=\small,auto]
\tikzstyle{arrow plain}=[arrs]
\tikzstyle{arrow dashed}=[dashed,arrs]
\tikzstyle{arrow bold}=[very thick,arrs]
\tikzstyle{arrow hide}=[draw=white!0,-]
\tikzstyle{arrow reverse}=[latex-]
\tikzstyle{cdnode}=[]

\definecolor{hexcolor0xa9a9a9}{rgb}{0.663,0.663,0.663}  
\tikzstyle{GrayLine}=[dashed,draw=hexcolor0xa9a9a9]
\tikzstyle{gray}=[dashed,draw=hexcolor0xa9a9a9]

\theoremstyle{definition}

\newtheorem*{theorem*}{Theorem}

\newtheorem{example*}[theorem]{Example*}
\newtheorem{examples*}[theorem]{Examples*}

\newtheorem{remark*}[theorem]{Remark*}

\def\bR{\begin{color}{red}}
\def\bB{\begin{color}{blue}}
\def\bM{\begin{color}{magenta}}
\def\bC{\begin{color}{cyan}}
\def\bW{\begin{color}{white}}
\def\bMl{\begin{color}{black}}
\def\bG{\begin{color}{green}}
\def\bY{\begin{color}{yellow}}
\def\e{\end{color}\xspace}
\newcommand{\bit}{\begin{itemize}}
\newcommand{\eit}{\end{itemize}\par\noindent}
\newcommand{\ben}{\begin{enumerate}}
\newcommand{\een}{\end{enumerate}\par\noindent}
\newcommand{\beq}{\begin{equation}}
\newcommand{\eeq}{\end{equation}\par\noindent}
\newcommand{\beqa}{\begin{eqnarray*}}
\newcommand{\eeqa}{\end{eqnarray*}\par\noindent}
\newcommand{\beqn}{\begin{eqnarray}}
\newcommand{\eeqn}{\end{eqnarray}\par\noindent}

\title{Foundations for Near-Term\\ Quantum Natural Language Processing}
\author{Bob Coecke, Giovanni de Felice, Konstantinos Meichanetzidis, Alexis Toumi\\ \ \\
Cambridge Quantum Computing Ltd.  \\ 
Oxford University, Department of Computer Science
}  

\begin{document}  
\maketitle 
 
\begin{abstract}        
We provide conceptual and mathematical foundations for near-term quantum natural language processing (QNLP), and do so  
in quantum computer scientist friendly terms.  We opted for an expository  presentation style, and provide references for supporting  empirical evidence  and formal statements concerning mathematical generality. 

We recall how the quantum model for natural language that we employ \cite{CSC} canonically combines linguistic meanings with rich linguistic structure, most notably grammar.
In particular, the fact that it takes a quantum-like model to combine meaning and structure, establishes QNLP as quantum-native, on par with simulation of quantum systems.  
Moreover, the now leading Noisy Intermediate-Scale Quantum (NISQ) paradigm for encoding classical data on quantum hardware, variational quantum circuits, makes NISQ exceptionally QNLP-friendly: linguistic structure can be encoded as a free lunch, in contrast to the apparently  exponentially expensive classical encoding of grammar.  

Quantum speed-up for QNLP tasks has already been established in previous work \cite{WillC}. Here we provide a broader range of tasks which all enjoy the same advantage. 

Diagrammatic reasoning is at the heart of QNLP.  Firstly, the quantum model interprets  language as quantum processes via the diagrammatic formalism of categorical quantum mechanics \cite{CKbook}. Secondly, these diagrams are via ZX-calculus translated into  quantum circuits. Parameterisations of meanings then become the circuit variables to be learned: 
\beq\label{eq:abs} 
\input{abs1.tikz}\ \ \ \mapsto\ \ \ \input{abs2.tikz}
\eeq

Our encoding of linguistic structure within quantum circuits also embodies a novel approach for establishing word-meanings that goes beyond the current standards in mainstream AI, by placing linguistic structure at the heart of Wittgenstein's meaning-is-context.   
\end{abstract}  


\newpage
\tableofcontents

\section{Introduction}     

We recently performed quantum natural language processing (QNLP) on IBM-hardware \cite{Nature}.   Given that we are still in the era of Noisy Intermediate Scale Quantum (NISQ) hardware, and that modern NLP is highly data intensive, the fact that we were able to do so may come as a bit of a surprise.   So what made this possible?  
 
It all boils down to the particular brand of NLP we used to underpin QNLP, which happens to not only be quantum-native, to not only be NISQ friendly, but in fact, it is NISQ that is QNLP friendly!   In particular, while encoding linguistic structure within our well-crafted NISQy setting comes as a free lunch,   doing so classically turns out to be very expensive, to the extend that standard methods now rather learn that structure than having it encoded.
 
Our  starting point is the DisCoCat brand of NLP \cite{CCS, CSC}, which constitutes a model of natural language that  combines linguistic meaning and linguistic structure such as grammar into one whole. While the model makes perfect sense without any reference to quantum theory, its true origin is the  categorical quantum mechanics (CQM) formalism \cite{AC1, kindergarten, CKbook}, as it was indeed the latter that enabled one to jointly accommodate linguistic meaning and linguistic structure.
Therefore,  it is natural to consider implementing this model on quantum hardware. Quantum advantage had already been identified in \cite{WillC} some years ago, and more recently, NISQ-implementation was considered in \cite{QPL-QNLP}. A first hardware implementation was reported in the blog-post \cite{QNLPmedium}.
 
An important feature of QNLP is  its critical reliance on diagrammatic reasoning.  We make use of no less than three diagrammatic theories, namely DisCoCat and CQM, both mentioned above, and also the versatile ZX-calculus \cite{CD1, CD2, CKbook}.  Firstly, by means of their common diagrammatic representation respectively in terms of DisCoCat and CQM, we can interpret  language as quantum processes.  Secondly, using ZX-calculus we morph the resulting quantum representation of natural language into quantum  circuit form, so that it becomes implementable on the existing quantum hardware.  This second use of diagrammatic theory goes well beyond `diagrams as convenient aids'. 

That said, we do not expect the reader to have any pre-existing knowledge on diagrammatic reasoning, as we provide all necessary background in quantum computer scientist friendly terms.
Also, the tensor network enthusiast  may find our presentations particularly appealing.  Overall, this paper is partly a review of relevant background material, partly provides some key new ideas, and brings all of these together in the form of some kind of manifest for NISQ-era quantum natural language processing. The key new ideas are:\vspace{-1mm}
\bit
\item the  translation of  linguistic structure into quantum circuits,\vspace{-1mm}
\item a NISQ-era pipeline for QNLP using variational quantum circuits, \vspace{-1mm}
\item and, a unified family of NLP-tasks that enjoy quantum speed-up.   
\eit

In Section \ref{sec;quantummodel} we recall our quantum model for natural language.  In our presentation we take noun-meanings to be qubit states, in order to provide natural language with a quantum computational `feel' right from the start, and all of the ingredients of our model then become well-known quantum computational concepts such as Bell-states.  In Section  \ref{sec;costadv} we discuss the origin and motivation for this quantum model, with particular focus on the fact that mathematically speaking it is canonical.  In fact, at the time,  the quantum model provided an answer to an important open problem in NLP.  We discuss the difficulties for its classical implementation, and its quantum advantage, including a new range of quantum computational tasks. In Section   \ref{sec;QNLPfriendly} we  present our implementation of QNLP on NISQ-hardware, with ZX-calculus playing the role of the translator which takes us from linguistic diagrams to quantum circuits, like in (\ref{eq:abs}).  In Section  \ref{sec:variationalCirc} we  explain how data can be `loaded' on the quantum computer using variational circuits, and how this paradigm makes NISQ exceptionally QNLP-friendly.  We also explain how our quantum implementations follow a novel learning paradigm that takes Wittgenstein's conception of meaning-is-context to a new level, that provides linguistic structure with the respect it deserves. 

\[
\epsfig{figure=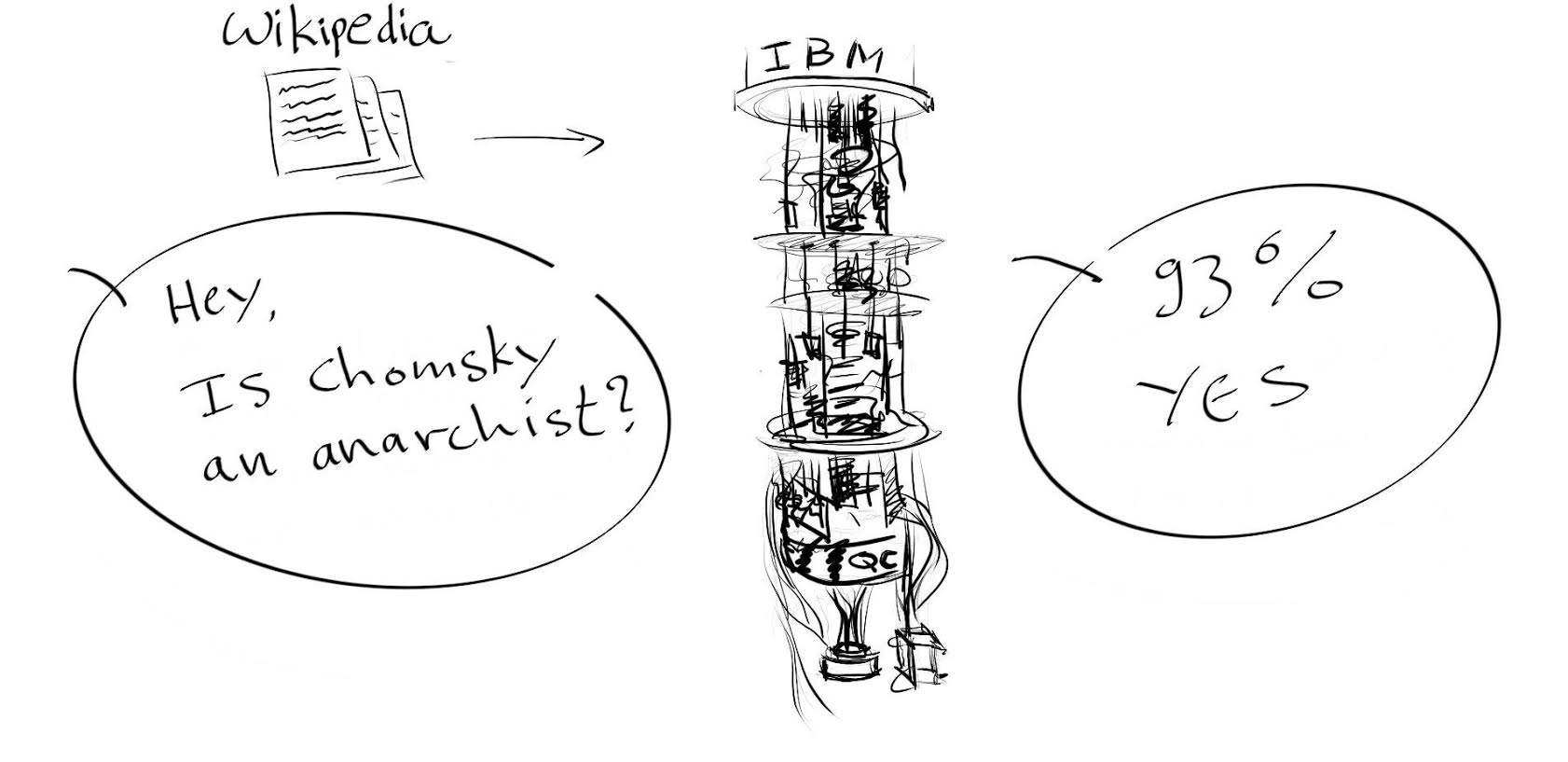,width=454pt} 
\]

\section{A quantum-model for natural language}\label{sec;quantummodel}    
 
We now present the model of natural language introduced in \cite{CSC}, recast in quantum computational terms. Assume for now that meanings of words can be described as states of some qubits, and in particular, that the meaning of a noun ${\tt n}$ like {\tt hat} is a 1-qubit state $|\psi_{\tt n}\rangle\in\mathbb{C}^2$.  
 
\subsection{Applying an adjective to a noun}\label{sec:applAdjN}  
 
An adjective ${\tt a}$ is something that modifies a noun, for example the adjective {\tt black} in {\tt black hat} modifies {\tt hat} to being {\tt black},  and therefore it is natural to represent it as a 1-qubit map $\eta_{\tt a}:\mathbb{C}^2\to\mathbb{C}^2$.   
The meaning of the adjective applied to the noun, say  $|\psi_{{\tt a}\cdot {\tt n}}\rangle\in\mathbb{C}^2$, is then: 
\[
|\psi_{{\tt a}\cdot {\tt n}}\rangle\ = \ \eta_{\tt a}\Bigl(|\psi_{\tt n}\rangle\Bigr)   
\]
Using diagrammatic notation like in \cite{kindergarten, ContPhys, CKbook}, which historically traces back to Penrose' diagrammatic calculus \cite{Penrose}, our example {\tt black hat} looks as follows:   
\beq\label{eq:adj1}
\input{an.tikz} 
\eeq 
where we read from top to bottom, and we made the following substitutions:  
\[
|\psi_{{\tt a}\cdot {\tt n}}\rangle\  \leadsto\ \begin{tikzpicture}[scale=0.80]
	\begin{pgfonlayer}{nodelayer}
		\node [style=none] (0) at (0, 0.25) {};
		\node [style=langstate] (1) at (0, 0.5) {\!\!\!{\tt black hat}\!\!\!};
		\node [style=none] (2) at (0, -0.75) {};
	\end{pgfonlayer}
	\begin{pgfonlayer}{edgelayer}
		\draw [style=swap] (0.center) to (2.center);
	\end{pgfonlayer}
\end{tikzpicture} 
\qquad\qquad
|\psi_{\tt n}\rangle\  \leadsto\ \begin{tikzpicture}[scale=0.80]
	\begin{pgfonlayer}{nodelayer}
		\node [style=none] (0) at (0, 0.25) {};
		\node [style=none] (1) at (0, -0.75) {};
		\node [style=langstate] (2) at (0, 0.5) {\!\!\!{\tt hat}\!\!\!};
	\end{pgfonlayer}
	\begin{pgfonlayer}{edgelayer}
		\draw [style=swap] (0.center) to (1.center); 
	\end{pgfonlayer}
\end{tikzpicture} 
\qquad\qquad
\eta_{\tt a}\  \leadsto\ \input{an3.tikz} 
\]

There is another manner for obtaining the meaning of {\tt black hat} from the meanings of {\tt black} and {\tt hat}, which allows one to also think of an adjective as a state, namely a 2-qubit state $|\psi_{\tt a}\rangle\in\mathbb{C}^2\otimes\mathbb{C}^2$, that is obtained from $\eta_{\tt a}$ via the Choi-Jamiolkowski correspondence.   Explicitly, in terms of matrix entries,  a $2\times 2$ matrix becomes a two qubit state as follows:
\[
\eta_{00} |0\rangle\langle 0| + \eta_{01} |0\rangle\langle 1| + \eta_{10} |1\rangle\langle 0| + \eta_{11} |1\rangle\langle 1| 
\ \ \mapsto \ \
\eta_{00} |00\rangle + \eta_{01} |01\rangle + \eta_{10} |10\rangle + \eta_{11} |11\rangle 
\]
Then, another way of writing the resulting state for applying an adjective to a noun is:
\beq\label{telesymb1}
|\psi_{{\tt a}\cdot {\tt n}}\rangle \ =\ \Bigl( \mathbb{I} \otimes \left\langle Bell\right| \Bigr)\circ\Bigl(|\psi_{\tt a}\rangle\otimes |\psi_{\tt n}\rangle\Bigr)
\eeq
where $\left\langle Bell\right|= \langle 00 | + \langle 11|$ and $\mathbb{I}$ is the identity. For our example, diagrammatically this gives:
\[
\input{anbis.tikz} 
\]
where now we made the substitution:  
\[
|\psi_{\tt a}\rangle\  \leadsto\ \input{an4.tikz} 
\]
One could symbolically compute that we indeed have: 
\[
\input{antris.tikz}  
\]
However, doing so  diagrammatically,  this becomes essentially a triviality.   Representing the Bell-state/effect by cap-/cup-shaped bend wires: 
\[
\left| Bell\right\rangle=\  \begin{tikzpicture}[scale=0.80]
	\begin{pgfonlayer}{nodelayer}
		\node [style=none] (0) at (-1.25, -0.5) {};
		\node [style=none] (1) at (1.25, -0.5) {};
	\end{pgfonlayer}
	\begin{pgfonlayer}{edgelayer}
		\draw [style=swap, bend left=90, looseness=1.50] (0.center) to (1.center);
	\end{pgfonlayer}
\end{tikzpicture} \qquad\qquad\qquad\left\langle Bell\right|=\  \begin{tikzpicture}[scale=0.80]
	\begin{pgfonlayer}{nodelayer}
		\node [style=none] (0) at (1.25, 0.5) {};
		\node [style=none] (1) at (-1.25, 0.5) {};
	\end{pgfonlayer}
	\begin{pgfonlayer}{edgelayer}
		\draw [style=swap, bend right=90, looseness=1.50] (1.center) to (0.center);
	\end{pgfonlayer}
\end{tikzpicture} 
\]
the Choi-Jamiolkowski correspondence becomes: 
\beq\label{eq:adj2}
\input{CJ.tikz}
\eeq
Hence, yanking the wire we obtain:
\[
\input{anquad.tikz} 
\]
Quantum computationally, this particular scheme is known as logic-gate teleportation \cite{Gottesman, LE}. 

The advantage of the form (\ref{telesymb1}) is that we now have the word meanings, all being states, tensorred together to give $|\psi_{\tt black}\rangle\otimes |\psi_{\tt hat}\rangle$, reflecting how we string words together to make bigger wholes like phrases and sentences. Then we apply the map $\mathbb{I} \otimes \left\langle Bell\right|$, which makes the meanings of the words interact, in order to produce the meaning of the whole.  As we shall see below, this map represents the grammar of the whole.

\subsection{Feeding a subject and an object into a verb}  

A transitive verb $tv$ is something that when provided with a subject ${\tt n}_s$ and an object ${\tt n}_o$ produces a sentence. For example, the verb {\tt hates} requires a subject, e.g.~{\tt Alice}, and an object, e.g.~{\tt Bob}, in order to form the meaningful whole {\tt Alice hates Bob}. Let's assume for now that such a sentence lives in $\mathbb{C}^{2k}$, where $k$ determines the size of the space where the meanings of sentences live.   Hence a transitive verb is represented as a map $\eta_{\tt tv}$ that takes two nouns $|\psi_{{\tt n}_s}\rangle\in\mathbb{C}^2$ and $|\psi_{{\tt n}_o}\rangle\in\mathbb{C}^2$ and produces $|\psi_{{\tt n}_s\!\cdot {\tt tv}\cdot {\tt n}_o}\rangle\in\mathbb{C}^{2k}$.  Using bilinearity of the tensor product we then have:
\[
|\psi_{{\tt n}_s\!\cdot {\tt tv}\cdot {\tt n}_o}\rangle = \eta_{\tt tv}\Bigl(|\psi_{{\tt n}_s}\rangle\otimes|\psi_{{\tt n}_o}\rangle\Bigr)
\]
For our example, diagrammatically we then have:
\[
\input{tv.tikz} 
\]
where the thick wire stands for $\mathbb{C}^{2k}$ and we made the substitutions:
\[
|\psi_{{\tt n}_s}\rangle\  \leadsto\ \begin{tikzpicture}[scale=0.80]
	\begin{pgfonlayer}{nodelayer}
		\node [style=none] (0) at (0, 0.25) {};
		\node [style=none] (1) at (0, -0.75) {};
		\node [style=langstate] (2) at (0, 0.5) {\!\!\!{\tt Alice}\!\!\!};
	\end{pgfonlayer}
	\begin{pgfonlayer}{edgelayer}
		\draw [style=swap] (0.center) to (1.center);
	\end{pgfonlayer}
\end{tikzpicture}
\qquad\qquad
|\psi_{{\tt n}_o}\rangle\  \leadsto\ \begin{tikzpicture}[scale=0.80]
	\begin{pgfonlayer}{nodelayer}
		\node [style=none] (0) at (0, 0.25) {};
		\node [style=none] (1) at (0, -0.75) {};
		\node [style=langstate] (2) at (0, 0.5) {\!\!\!{\tt Bob}\!\!\!};
	\end{pgfonlayer}
	\begin{pgfonlayer}{edgelayer}
		\draw [style=swap] (0.center) to (1.center);
	\end{pgfonlayer}
\end{tikzpicture}
\qquad\qquad
\eta_{\tt tv}\  \leadsto\ \input{tv1.tikz}
\]

Using a more funky variant of the Choi-Jamiolkowski correspondence we can also represent  the transitive verb as a state $|\psi_{\tt tv}\rangle\in\mathbb{C}^2\otimes
\left(\mathbb{C}^{2k}\right)\otimes\mathbb{C}^2$. This can again be best described using the diagrammatic representation:
\[
\input{CJhates.tikz}
\]
We can now compute the meaning of a sentence as follows:
\[
\input{tv2.tikz} 
\]
Back to Dirac-notation, this is:
\beq\label{telesymb2}
|\psi_{{\tt n}_s\!\cdot {\tt tv}\cdot {\tt n}_o}\rangle \ =\ 
\Bigl(
\left\langle Bell\right| \otimes \mathbb{I}\otimes \left\langle Bell\right|
\Bigr)
\circ
\Bigl(
|\psi_{{\tt n}_s}\rangle\otimes |\psi_{\tt tv}\rangle  \otimes|\psi_{{\tt n}_o}\rangle
\Bigr)
\eeq
Quantum computationally, this is a variant of logic-gate teleportation where now two states are made to interact by means of a gate that is encoded as a 3-system state.   

Again, just like in the previous section, in the shape (\ref{telesymb2}) we first tensor all word meanings together, in this case $|\psi_{\tt Alice}\rangle\otimes |\psi_{\tt hates}\rangle\otimes |\psi_{\tt Bob}\rangle$, and then
we apply the map $\left\langle Bell\right| \otimes \mathbb{I}\otimes \left\langle Bell\right|$, which represents the grammar.

\subsection{Functional words such as relative pronouns}   

\paragraph{Spiders.}  Below we will  make use of spiders \cite{CPV, CKbook}.  These spiders are also the key ingredient of the ZX-calculus \cite{CD1, CD2}.   Concretely, spiders are the following linear maps, defined relative to a given orthonormal basis $\{ |i\rangle \}_i$:
\[ 
\input{spidercomp.tikz}\ \ = \ \ \sum_i |i\ldots i\rangle\langle i\ldots i|  
\] 
Some examples of spiders are `copy', `merge' and `delete':
\[
\input{copy.tikz}\ \ =\ \  \sum_i |i i\rangle\langle i|\qquad\qquad\input{merge.tikz}\ \ = \ \ \sum_i |i\rangle\langle i i|
\qquad\qquad\begin{tikzpicture}[scale=0.80]
	\begin{pgfonlayer}{nodelayer}
		\node [style=none] (0) at (0, 0.75) {};
		\node [style=white dot] (1) at (0, -0.5) {};
		\node [style=none] (2) at (0, -0.5) {};
	\end{pgfonlayer}
	\begin{pgfonlayer}{edgelayer}
		\draw [style=swap, in=90, out=-90, looseness=0.75] (0.center) to (1); 
	\end{pgfonlayer}
\end{tikzpicture}\ \ =\ \  \sum_i \langle i|
\]
Identities and the caps and cups that we saw above are also special cases of spiders:  
\[
\begin{tikzpicture}[scale=0.80]
	\begin{pgfonlayer}{nodelayer}
		\node [style=none] (0) at (0, -1) {};
		\node [style=none] (1) at (0, 1) {};
	\end{pgfonlayer}
	\begin{pgfonlayer}{edgelayer}
		\draw [style=swap] (0.center) to (1.center);
	\end{pgfonlayer}
\end{tikzpicture}\ \ =\ \  \sum_i |i \rangle\langle i |
\qquad\qquad
\begin{tikzpicture}[scale=0.80]
	\begin{pgfonlayer}{nodelayer}
		\node [style=none] (0) at (1.25, -0.5) {};
		\node [style=none] (1) at (-1.25, -0.5) {};
	\end{pgfonlayer}
	\begin{pgfonlayer}{edgelayer}
		\draw [style=swap, bend left=90, looseness=1.50] (1.center) to (0.center); 
	\end{pgfonlayer}
\end{tikzpicture}\ \ =\ \  \sum_i |i i\rangle
\qquad\qquad
\begin{tikzpicture}[scale=0.80]
	\begin{pgfonlayer}{nodelayer}
		\node [style=none] (0) at (1.25, 0.5) {};
		\node [style=none] (1) at (-1.25, 0.5) {};
	\end{pgfonlayer}
	\begin{pgfonlayer}{edgelayer}
		\draw [style=swap, bend right=90, looseness=1.50] (1.center) to (0.center);
	\end{pgfonlayer}
\end{tikzpicture}\ \ = \ \ \sum_i \langle i i|
\]
One can easily show that spiders are subject to the following fusion rule:
\[ 
\input{spider.tikz}\ \ =\ \ \input{spidercomp.tikz}
\] 
that is, if we compose spiders together, however many and in whatever way, the result is always a spider.  In many cases this rule will also be applied in the opposite direction, namely, un-fusing a single spider into multiple spiders. All together, the most intuitive way to think about these spiders is that all that matters is what is connected to what by means of spiders, and not the manner in which these connections are established.  

\par\bigskip

In general, special words will be represented by special states. For example, following \cite{FrobMeanI, FrobMeanII}, relative pronouns {\tt rp} like {\tt who} are represented as follows:
\beq\label{eq:DiracBad}
|\psi_{\tt who}\rangle = \Bigl(|00\rangle \Bigl(\sum_i |i\rangle \Bigr) |0\rangle\Bigr) + \Bigl(|11\rangle \Bigl(\sum_j |j\rangle \Bigr) |1\rangle\Bigr)
\eeq
If we only consider the 1st, 2nd and 4th qubit, we have the GHZ-state:
\[
\left|GHZ\right\rangle \ = \ |000\rangle + |111\rangle
\]
and if we only consider the 3rd qubit have the higher-dimensional $|+\rangle$-state.  So what we have here is the higher-dimensional $|+\rangle$-state in between the 2nd and the 3rd qubit of a GHZ-state. Each of these are in fact spiders, so diagrammatically the picture becomes much clearer: 
\[
\input{rp.tikz} 
\]
We can now compute the meaning of noun-phrases involving relative pronouns as follows:
\beq\label{eq:relpronQNLP}
\input{relpronQNLP.tikz} 
\eeq
As should already be clear from (\ref{eq:DiracBad}), at this point it  becomes increasingly difficult to represent computations like (\ref{eq:relpronQNLP}) still in Dirac-notation.  We can just about still  do it, when first simplifying the wirings a bit in order to obtain:
\beq\label{eq:relpronQNLP2} 
\input{relpronQNLP2.tikz}  
\eeq
So symbolically we obtain: 
\beq\label{telesymb3}
|\psi_{{\tt n}_{h}\!\cdot {\tt rp}\cdot {\tt tv}\cdot {\tt n}_o}\rangle \ =\
\Bigl(
\bigl( |0\rangle\langle 00|+|1\rangle\langle 11| \bigr) \otimes \Bigl(\sum_{j=1}^{j=k} \langle j| \Bigr)\otimes \left\langle Bell\right|
\Bigr)
\circ
\Bigl(
|\psi_{{\tt n}_{h}}\rangle\otimes |\psi_{\tt tv}\rangle  \otimes|\psi_{{\tt n}_o}\rangle
\Bigr)
\eeq
In quantum computational terms, here we are using a fusion operation \cite{browne2005resource} and a (coherent) deletion operation.  In fact, diagrams like (\ref{eq:relpronQNLP}) and (\ref{eq:relpronQNLP2}) can be directly implemented within quantum optics, hence making optics `do language' just as humans do. This shouldn't come as a surprise, as it was also quantum optics that was used to do the first implementations of protocols like quantum teleportation \cite{TeleExp} and entanglement swapping \cite{Swap}, and it were these kinds of protocols that provided inspiration for our quantum model of language \cite{teleling}.

\subsection{The general case}  
 
The quantum model for language meaning introduced above---although not for the specific case of qubits---is what we referred to in the introduction as DisCoCat.  In particular, equations (\ref{telesymb1}), (\ref{telesymb2}) and (\ref{telesymb3}) are instances of a general algorithm, introduced in \cite{CSC}, that allows one to produce the meaning of a whole from its constituent words. We've already seen examples of doing so for a sentence and a noun-phrase.  We now describe this general algorithm.

Assume as given a string of words ${\tt w}_1\cdot\ldots\cdot{\tt w}_N$ that forms the `whole', and that each word comes with a representation $|\psi_{{\tt w}_i}\rangle\in\mathbb{C}^{k_i2}$ where the dimension of this space depends on the grammatical type. For example, dimension $2$ for a noun, dimension $2\times 2$ for an adjective etc.---you learn the relationship between the dimensions of different grammatical types from a grammar book like \cite{LambekBook}.  We then tensor these states together: 
\[
|\psi_{{\tt w}_1}\rangle\otimes\ldots\otimes|\psi_{{\tt w}_N}\rangle
\]
In this representation,  all the words are completely disentangled, using a term from quantum theory.  Pictorially, this means that they are disconnected:
\[
\input{tv2ng.tikz}
\]
So in particular,  there is no interaction between  any of the words.  Now using a term from NLP \cite{harris1954distributional}, they merely  form  a bag-of-words.
 
What makes these words interact is the grammatical structure, or in other terms, the grammatical structure `binds these words together, in order to form an entangled whole. So will now let grammar enter the picture. Consulting again a grammar book like \cite{LambekBook}, given the grammatical structure of the sentence ${\cal G}$, we can now construct a map $f_{\cal G}$ like:
\[
\left\langle Bell\right| \otimes \mathbb{I}\otimes \left\langle Bell\right|
\]
which we encountered in equation 
(\ref{telesymb2}), in a picture:
\[
\input{tv2nm.tikz}
\]
Such a map can always be made using only Bell-effects and identities, like in all of the examples that we have seen.  The reason for this is that grammar as defined in the book \cite{LambekBook} is an algebraic gadget called pregroups, that is all about having cup-shaped wires.  We say a bit more about these algebraic gadgets in the next section.

Then, using this map $f_{\cal G}$,  we unleash grammatical structure on the string of words: 
\[
f_{\cal G}\Bigl(|\psi_{{\tt w}_1}\rangle\otimes\ldots\otimes|\psi_{{\tt w}_N}\rangle\Bigr)  
\]  
This results in the meaning of the whole, in a picture:   
\[
\input{tv2wg.tikz} 
\]

At this point, diagrammatic notation is becoming unavoidable.  In particular, as sentences grow, denoting the map $f_{\cal G}$ in Dirac-notation becomes practically impossible.  For example, try to imagine how the map $f_{\cal G}$ in the following example would look like in Dirac-notation: 
\beq\label{eq:BIG} 
\input{flowers2bis.tikz}
\eeq
In fact, the wire-representation does quite a bit more besides merely providing denotational convenience.  Firstly, it tells us how meanings `flow' between the words.  For example, here the wires show how the subject- and object-meanings flow into the verb: 
\[
\input{tv2wgf.tikz} 
\]
Hence, they provides us with a very intuitive  understanding of grammar:  
\begin{center}
\fbox{\em Grammar is what mediates the flows of meanings between words.\em}   
\end{center}
This is indeed also how we derived the quantum representations for adjectives and transitive verbs above, with an adjective waiting for a noun's meaning and modify it, and a transitive verb waiting for a subject's and an object's meaning, and making these interact. 

Secondly, once we adopt this idea of flows of meanings between words, we can reverse-engineer the meanings of words themselves, like in the case of {\tt that}, {\tt does} and {\tt not}.  In the case of {\tt does}, the wires simply propagate the meaning of the subject without altering them \cite{CSC}. In the case of {\tt not}, in addition to propagating the meaning, negation is applied to the resulting sentence \cite{CSC}.  In the case of {\tt that}, the noun {\tt flowers} gets conjoined---that is to say, {\tt and}---resulting phrase should be a noun-phrase \cite{FrobMeanI, CoeckeText}.  

Something else that we can easily see from the diagrammatic representations is that there is plenty of entanglement.  This entanglement increases even more when one moves to larger text. In text specific nouns may occur throughout the entire text, and this will cause multiple sentences to become entangled.  This representation for larger text, which opens up an entirely new world, can be found in \cite{CoeckeText}.  In this new world,  meanings are not static anymore, but evolve as text progresses.  In the present paper we won't go in any great detail of text representation.  In Section \ref{sec:complangcirc} we do indicate how  sentences can be composed to form larger texts.
 
\subsection{Comparing meanings}  

Now that we know how to compute meanings of multi-word composites such as phrases and sentences, we now want to compare these composites. More specifically, we want to investigate if different sentences may convey closely related meanings. For example, we want to figure out whether the following two are closely related:
\begin{center}
{\tt Alice hates Bob}\qquad and\qquad{\tt Alice does not like Bob}
\end{center} 
In order to do so, we can now use the inner-product.  Given two meanings $|\psi_{{\tt w}_1\cdot\ldots\cdot{\tt w}_N}\rangle$ and $|\psi_{{\tt w}_1'\cdot\ldots\cdot{\tt w}_N'}\rangle$, as always, we turn one of these kets into a bra, say: 
\[
|\psi_{{\tt w}_1\cdot\ldots\cdot{\tt w}_N}\rangle\ \mapsto \ \langle\psi_{{\tt w}_1\cdot\ldots\cdot{\tt w}_N}|
\] 
and then we form the bra-ket:
\[
\langle\psi_{{\tt w}_1\cdot\ldots\cdot{\tt w}_N}|\psi_{{\tt w}_1'\cdot\ldots\cdot{\tt w}_N'}\rangle
\]
Then, the number $|\langle\psi_{{\tt w}_1\cdot\ldots\cdot{\tt w}_N}|\psi_{{\tt w}_1'\cdot\ldots\cdot{\tt w}_N'}\rangle|$, or the number $|\langle\psi_{{\tt w}_1\cdot\ldots\cdot{\tt w}_N}|\psi_{{\tt w}_1'\cdot\ldots\cdot{\tt w}_N'}\rangle|^2$, 
or any monotone function thereof, is a measure of similarity between the two sentences.  The same can be done in diagrammatic notation, for example, first we turn a ket into a bra:
\[
\input{tvket.tikz}\ \ \mapsto\ \ \input{tvbra.tikz}
\]
and then we form the bra-ket: 
\[
\input{tvbraket.tikz}
\]

In order to form such an inner-product, meanings need to have the same dimension.  This is for example the case for an individual noun and any noun-phrase, and one expects: 
\[ 
\input{queen.tikz}
\] 
to be close to maximal.  The key point here is that however sentences, noun phrases, or any other whole are made up (i.e.~what their grammatical structure is), as long as the resulting grammatical type is the same, they can be compared.  On the other hand, a noun and a sentence are very different linguistic entities and cannot be straightforwardly compared.    

\section{Why quantum?}\label{sec;costadv} 

How did this model come about, and how does it perform? 

\subsection{Combining grammar and meaning}\label{sec:combiningmeaninggrammar}  
 
The quantum model for natural language certainly did not came about because of some metaphysical, spiritual, salesman, PR or any other crackpot motivation that aimed to link natural language with quantum physics.  In fact, when it came out, the authors Coecke, Sadrzadeh and Clark were denying  any reference to quantum theory in order to avoid any such connotation.\footnote{Of course, unavoidably, the popular media didn't let them get away with it, with ``quantum linguistics" making the cover of New Scientist \cite{NewScientist1}, Coecke being branded a ``quantum linguist" by FQXi \cite{FQXi}, and in Scientific American DisCoCat being being referred to as  ``quantum mechanical words" \cite{ScAm}. Of course, in the end, all were of course very grateful for that level of attention.}  

The true origin of this model was a purely practical concern, namely, it provided an answer to an important open question in the field of NLP: If we know the meanings of words, can we compute the meaning of a sentence that is made up of these words?  By then, NLP had been very successful  for several important tasks like automatically establishing synonyms, which it did by computing inner-products. The aim was then to extend this method to phrases and sentences,  just as we did in the previous section.  And as we  discussed in that section, this would  require a representation of sentence meanings that would be independent of the grammatical structure.  Such a representation didn't exist yet before  DisCoCat was proposed.     
 
This specific problem had been stated by Clark and Pulman \cite{ClarkPulman}, who also pointed at Smolensky's representation in the neuroscience literature that combines meaning and grammar \cite{Smolensky, SmolenskyBook}.\footnote{Recently, Smolensky's model also showed up in the quantum computing literature \cite{Wiebe-Smolensky}. Here,  simulated annealing is used to compute the optimal (i.e.~maximum `harmony') parse tree for a given sentence, and they show this problem is BQP-complete.  Somewhat related, in  \cite{bausch2019quantum} the authors use the quantum maximum finding algorithm, a variant of Grover's quantum search, in order to obtain superquadratic speedups for the problem of finding the best parse in a context-free language.} However, Smolensky's representation space was highly dependent on the grammatical structure, and in particular, it grew with the size of the sentence.  A higher-level aim stated in \cite{ClarkPulman} was to bring structural aspects of language (like grammar theory) and empirical/statistical approaches (these days mostly machine learning) back together \cite{Gazdar}, and also today, even after the amazing successes of deep learning, that quest is still on---see e.g.~\cite{pearl2018book, chollet2019measure}.   

\paragraph{Grammar.} The mathematics of grammar goes back as far as the first half of the previous century \cite{Ajdukiewicz, Bar-Hillel}.  Lambek's seminal work is now still being used \cite{Lambek0}, and that era was also marked by  Chomsky's important contributions \cite{Chomsky, ChomskyBook1, ChomskyBook2}. A change of mind by Lambek at the very end of the previous century \cite{Lambek1} resulted in a new formalism for grammar, namely pregroups, which on-the-nose correspond to these wire structures:
\beq\label{DisCo1}
\input{grammar.tikz}   
\eeq
Here, these cup-shaped wires have nothing to do with Bell-effects, they are nothing but wires.  Initially they were not presented as such.  Instead, they were presented within the context of a weakening of the notion of a group, called a pregroup. Rather than wires, one then encounters calculations that look as follows:
\beq\label{pregroupcalc}
n   \cdot   \left({}^{-1}n \cdot s \cdot n^{-1}\right)  \cdot n
\leq \left(n   \cdot   {}^{-1}n\right) \cdot s \cdot \left(n^{-1}  \cdot n\right) 
\leq  1 \cdot s\cdot 1 \leq  s 
\eeq
When moving to the realm of category theory, the nested wire-structures and pregroups can be shown to be equivalent. For this equivalence we refer the reader for example to \cite{Gospel}.

\paragraph{Meaning.} While grammar corresponds to plain wires, meaning has in recent decades been described by vectors representing empirical data \cite{jones1997readings}.  There are a number of different manners how these can be obtained, e.g.~co-occurence counts  \cite{harris1954distributional}, vector embeddings such as word2vec \cite{mikolov2013efficient} and Glove \cite{pennington2014glove}, and more recently the contextualised vector embeddings of Bert \cite{devlin2018bert}, and now GPT-3 \cite{brown2020language} is taking the centre stage.  So it is perfectly natural to use the same notation as we use for quantum states:
\beq\label{DisCo2}  
\input{meanings.tikz}  
\eeq
There are however some key differences with the representation of meanings as vectors in the standard NLP literature, and how they occur in the DisCoCat model.  

Firstly, in the standard NLP literature there  is, of course, nothing like `word-meanings made from wires' like the ones that we encountered in  examples (\ref{eq:relpronQNLP}) and (\ref{eq:BIG}) above: 
\beq\label{eq:intwire}
\input{meaningsaswires.tikz}
\eeq
When above we made reference to `linguistic structure such as grammar', the use of `linguistic structure' accounted also for these other uses of wires, besides grammar.  

Secondly, in most representations of meanings as vectors in the standard NLP literature everything lives in the same space, so rather than vectors looking like this: 
\[
\input{ManySpace.tikz}  
\]
they rather look like this:     
\[
\input{OneSpace.tikz}
\]
In Dirac notation, rather than:
\[
|\psi_{\tt Bob}\rangle\in\mathbb{C}^2 
\qquad\qquad
|\psi_{\tt black}\rangle\in\mathbb{C}^2 \otimes\mathbb{C}^2
\qquad\qquad
|\psi_{\tt hates}\rangle\in\mathbb{C}^2\otimes\mathbb{C}^{2k}\otimes\mathbb{C}^2
\]
we have:
\beq\label{meaningvectors}
|\psi_{\tt Bob}\rangle\ , \ |\psi_{\tt black}\rangle\ , \ |\psi_{\tt hates}\rangle \in\mathbb{C}^2 
\eeq

\paragraph{Grammar\,$\otimes$\,Meaning.} Now, given grammar calculations like (\ref{pregroupcalc}) and meaning vectors coming in the form (\ref{meaningvectors}), figuring out  that they get unified in pictures like (\ref{eq:BIG}) may seem to be an incredible tour de force.  Indeed, and this wouldn't have happened if it wasn't for a very big coincidence. Firstly, there was the fact that these pictures were already around in the form of CQM.   Secondly, within the same department, and even in the same corridor, there  were the authors of the paper \cite{ClarkPulman} where the question was posed. Thirdly, the authors of this paper knew vector representation (\ref{meaningvectors}) of course very well.  Fourth, again in the same corridor there was also the logician Sadrzadeh who knew the grammar theory calculations (\ref{pregroupcalc}) very well, having worked with Lambek himself on the pregroup structure of Persian \cite{SadrPersian}.  Hence, it ended up being a matter of identifying both grammar and meaning within those existing CQM pictures, which was pretty straightforward, as clearly indicated in (\ref{DisCo1}) and (\ref{DisCo2}).\footnote{To be strictly factual, the identification didn't happen in terms of pictures, but in terms of symbolic category theory (cf.~\cite{CSC}), which ultimately is the same thing anyway \cite{JS, SelingerSurvey, CatsII}.  In  categorical language, one easily observes that on the one hand pregroups,  and on the other hand the category in which vector spaces and linear maps organise themselves, have exactly the same structure, allowing one to faithfully embed grammar into the latter. It is indeed this connection between these two categories which motivates the instantiation of grammatical reductions as quantum processes, since quantum processes can be described in terms of vector spaces and linear maps, as well.
This is now the common view, as put forward in \cite{CSC}, although the initial categorical construction consisted pairing the two categories \cite{CCS}.} 

\paragraph{The `plain' DisCoCat algorithm.}  So here is a summary of how the algorithm that turns the meanings of parts in the meaning of a whole works.  
\ben
\item Assumed is that for a whole ${\tt w}_1\cdot\ldots\cdot{\tt w}_N$, which grammatically makes sense, the meanings of the parts are given, and  the spaces in which these meanings of parts live (cf.~$|\psi_{{\tt w}_i}\rangle\in\mathbb{C}^{k_i2}$) respect the grammatical type  according to \cite{LambekBook}, for example, we have:
\[
\input{Alg1.tikz}
\]
\item Using \cite{LambekBook}, we establish the wire diagram representing grammar, for example: 
\[
\input{Alg2.tikz}
\]
\item We replace all the cups in the diagrams with Bell-effects (of the right dimension) and all the straight wires with identities, as such producing a quantum process:
\[
\input{Alg3.tikz}
\]
\item We apply this quantum process to the meanings of the parts: 
\[
\input{Alg4.tikz}
\]
The result is the meaning of the whole. 
\een
  
\paragraph{The `augmented' DisCoCat algorithm.} The algorithm presented above only accounts for grammatical structure.  It can be augmented to account for more linguistic structure, most notably, using `meanings as wires' as in (\ref{eq:intwire}).  We will see some more examples of this below.
  
\bigskip\noindent
What we presented here was the original DisCoCat algorithm. However, as we will see further, we won't retain some of the basic principles presented here:  
\bit
\item Firstly, the bottom-up approach from meanings of the parts to the meaning of the whole won't be retained as there is no obvious manner for implementing that on NISQ hardware, nor is it competitive with current NLP platforms.\footnote{Presently this means platforms such a BERT and  GPT-3 but surely, in the not too far distant future, probably nobody would refer to these platforms anymore as they will have been taken over by others.}  
\item Secondly, the shape of the diagrams as presented here also won't be retained as there is no direct efficient implementation of them on NISQ hardware, nor do they extend to larger text that goes beyond single sentences.   
\eit 
As we will discuss below, neither of these departures constitutes a compromise, but instead represents genuine progress for the DisCoCat programme. 
  
\subsection{Quantum $\bigcap$ Natural Language = interaction logic}\label{sec:interaction}

So let's get back to the question: Why quantum?  It just happened to be the case that the quantum pictures provided the perfect blend of grammar and meaning, nothing more, nothing less.  We therefore preferred to use the term quantum-inspired \cite{teleling}.

In particular, there is no deep mystical connection between quantum theory and natural language.  There is something in common though, as indicated by the intersection symbol $\bigcap$ in the heading of this section. What they share is what we prefer to call an  `interaction logic' \cite{DBLP:books/daglib/p/Coecke17}.  On the one hand, it is fair to say that this interaction is what should have been called quantum logic, instead of the now forgotten field started by Birkhoff and von Neumann \cite{BvN}. On the other hand, this interaction logic is not even specific to natural language, but also governs many other things, for example, the manner in which our sensory inputs interact \cite{ConcSpacI}, and  how images and movies can be interpreted as interacting processes \cite{CoeckeText}.  

But this shared interaction logic by no means points at the fact that the descriptions of quantum theory and natural language are close to being the same. As a matter of fact, while it is mostly assumed that Hilbert spaces do play a fundamental role for quantum theory, for language they essentially represent statistics, and by no means constitute the foundational structure of language meaning. Indeed, as language can be used to talk about anything we know, it should then also encompass anything we know, and in particular, all the structures we use to describe anything we know.    Whatever that is...

This interaction logic is a very novel feature that in most sciences has never been identified until recently.\footnote{Of course, much of modern physics is all about interaction, for example, condensed matter physics.  What we are talking about here are logical systems, in contrast to traditional propositional logic which by definition focusses on the properties of a sufficiently isolated system.   In biological terms, the distinction constitutes describing an animal in terms of its genetical code, versus describing it by its behaviour in the wild \cite{ContPhys, CatsII}.  Going even further, what we are talking about here are logical systems as complete substitutes for more traditional formalisms.  These traditional formalisms are always bottom-up, first describing individual systems, then how to compose these.  We are talking about logical systems that start by describing interactions, and end by describing interactions.  ZX-calculus is an example of this, and has meanwhile shown to be complete for equational reasoning relative to the Hilbert space formalism \cite{hadzihasanovic2018two}.  That is, any equation that can be derived in Hilbert space can also be derived in ZX-calculus.}  The reason being that exact sciences are still highly reductionist, describing things by their make-up, rather than focussing on interactions of systems.  In the case of language it is clear that words interact tremendously, as witnessed by the wire structures in our diagrams.  Similarly, the same diagrams appear when we represent quantum teleportation \cite{LE, Kauffman, kindergarten}. As Schr\"odinger pointed out, in quantum theory we simply can't ignore the particular nature of how systems interact, and he even called it the defining feature of quantum theory \cite{SchrodingerComp}.
We think that it is fair to say that CQM is the logical incarnation of Schr\"odinger view, by taking interaction as the primal connective \cite{QLog}.     

\subsection{Classical successes, but costly}\label{sec:class}     

Soon after the development of the quantum model of language it was put to the test, on classical hardware of course, and the model greatly outperformed all other available models for a number of standard NLP tasks \cite{GrefSadr, KartSadr}. Despite the growing dominance of machine learning,  successes have continued until recently \cite{wijnholds-sadrzadeh-2019-evaluating, MarthaNeural, wijnholds2020representation}.  With the 1st DisCoCat paper appearing in 2008 in a somewhat obscure venue \cite{CCS}, independently Baroni and Zamparelli also proposed the adjective-noun model of Section \ref{sec:applAdjN}, again  strongly supported by experimental evidence \cite{baroni2010nouns}.

These experiments had one major obstacle: the reliance of the quantum model on the tensor product caused an exponential blow-up for the spaces in which meanings of words live.
Consider for example transitive verbs:  
\beq\label{eq:SentenceSpace}
\input{SentenceSpace.tikz} 
\eeq
Thus far we assumed meaning spaces to be low-dimensional, e.g.~nouns being represented by qubits.  But in practical NLP these dimensions can go up to a thousand, and even to a million.  Conservatively assuming that the thick wire is the same as two thin wires, that would take the dimension to $1,000,000^{(1+2+1)}=1,000,000^4=10^{24}$, which obviously wouldn't fit on any existing computer hardware.

\paragraph{A hack?} In order to avoid  this exponential blow-up, it was assumed that the verb-state (\ref{eq:SentenceSpace}) also had a certain internal wiring, namely the following one \cite{GrefSadr, KartsaklisSadrzadeh2014}:
\beq\label{eq:verb}
\input{verb.tikz} 
\eeq
So we start with a two-wire state {\tt *hates*}.  We then copy each of the wires, and as the picture also  indicates, we then bundle two of the resulting wires together in order to make up the thick wire, hence obtaining {\tt hates}.  In concrete Hilbert space terms, instead of having transitive verbs being described by states living in the large space:
\[
|\psi_{\tt hates}\rangle\in\mathbb{C}^2\otimes\mathbb{C}^{2k}\otimes\mathbb{C}^2
\]
we start with states living in a smaller space:
\[
|\psi_{\tt *hates*}\rangle\in\mathbb{C}^2\otimes\mathbb{C}^2
\]
and then lift these to the larger space as follows:
\[
\Bigl( \sum_i |ii\rangle\langle i| \otimes \sum_j |jj\rangle\langle j|  \Bigr)\Bigl(|\psi_{\tt *hates*}\rangle\Bigr) \in\mathbb{C}^2\otimes\mathbb{C}^{2\times 2}\otimes\mathbb{C}^2
\]
So the space reduction for the passage:
\[
\input{SentenceSpace.tikz}\ \ \mapsto\ \ \input{SentenceSpace2.tikz}
\]
is then, from $k+2$ thin wires to $2$ thin wires.  For $k=2$, then $1,000,000^{(1+2+1)}=10^{24}$ dimensions reduce to $1,000,000^{(1+1)}=10^{12}$ dimensions, and for more modest 1000 dimensional spaces, this becomes a very manageable $10^{6}$.

\paragraph{A feature!} In fact, this dimensional reduction via an internal wiring  is not just a hack, but conceptually speaking, makes  a lot of sense as well.  Bringing back {\tt Alice} and {\tt Bob} into the picture, we can now substitute the form (\ref{eq:verb}) in the sentence {\tt Alice hates Bob} and obtain: 
\[  
\input{tv3.tikz} 
\]
The picture now indicates that the verb imposes a certain relationship on Alice and Bob:  
\[  
\input{tv4.tikz} 
\]
namely the relationship {\tt hates}, and this is indeed what the sentence intends to communicate.   This case of transitive verbs is then to  be viewed as the entangled counterpart to the case where we impose an adjective on a noun:
\[  
\input{adjdot1.tikz} 
\]
At first it may seem that this does not exactly match our earlier discussion of adjectives as in  (\ref{eq:adj1}).  In fact, it does match, in the sense that it is again a special case, namely, the special case where  we assume that an adjective like {\tt black} also has an internal wiring: 
\[  
\input{adjdot2.tikz} 
\]
Or equivalently, when thinking of an adjective as a state as in (\ref{eq:adj2}), it is the special case: 
\[  
\input{adjdot3.tikz}  
\]
From these internal wirings, we obtain a new perspective on grammar, in terms of a mechanism for imposing properties on nouns \cite{CoeckeText, CoeckeMeich}.  For example,  adjectives and transitive verbs respectively impose unary properties and binary properties on nouns, while ditransitive verbs impose ternary properties on nouns.

\paragraph{Density matrices as meanings.} Following the initial experimental successes and positive reception, the quantum model for language was extended to  include more features of language.  To achieve that even more quantum structures were used, most notably density matrices.  These serve two distinct purposes within the context of modelling language: 
\bit

\item Firstly, as pointed out by Kartsaklis and Piedeleu in their respective theses, density matrices can serve the same role  as the one for which von Neumann introduced them for quantum theory, namely, to account for ambiguity \cite{DimitriDPhil, RobinMSc, calco2015}.  In quantum theory a mixed state can indeed reflect ambiguity about the precise pure state a system is in, and similarly, in language, a mixed state can reflect ambiguity about the exact meaning of a word.  Language is  full of ambiguities, and ambiguous words in particular, for example the word {\tt queen}, which can be a royal, a rock band, a bee, a chess piece, a playing card, etc.  In order to account for each of these meanings we can form the mixed state:
\beqa
\rho_{\mbox{\scriptsize\tt queen}} &=& |\mbox{\tt queen-royal}\rangle\langle\mbox{\tt queen-royal}|   \\
&+&|\mbox{\tt queen-band}\rangle\langle\mbox{\tt queen-band}|   \\  
&+&|\mbox{\tt queen-bee}\rangle\langle\mbox{\tt queen-bee}|   \\  
&+&|\mbox{\tt queen-chess}\rangle\langle\mbox{\tt queen-chess}|   \\  
&+&|\mbox{\tt queen-card}\rangle\langle\mbox{\tt queen-card}| \\
&+& \ldots 
\eeqa 
and  apply some normalisation if required.  Experiments using this representations are reported on in \cite{calco2015}.  It is shown that the mixedness of such a state, when put in an appropriate context, will not carry over to the larger whole provided the latter provides enough context for full disambiguation. For example,  the noun-phrase {\tt Freddy Mercury's queen} clearly singles out $|\mbox{\tt queen-band}\rangle\langle\mbox{\tt queen-band}|$ as the intended meaning.  In order to get a  feel for how this could work mathematically, we can represent the map {\tt Freddy Mercury's}---this is indeed a map that transforms a noun into a noun---as a projector $P\circ --\circ P$ acting on density matrices.  Assuming that the projector $P$ has {\tt queen-band} as a fixed-point and maps all other meanings of {\tt queen} to the zero-vector, we indeed obtain a pure state:  
\[
\input{queen1.tikz}\ \ = \ P \circ \rho_{\mbox{\scriptsize\tt queen}} \circ P\  = \ |\mbox{\tt queen-band}\rangle\langle\mbox{\tt queen-band}|
\]
While this is of course a hand-crafted idealised example, the experiments mentioned above show that something similar happens with real-world data. 

\item Secondly, there also are  hierarchical relationships between words, for example {\tt lion} is an example of {\tt big cat}, which is an example of {\tt mammal}, which is an example of {\tt vertebrate} etc.  While normalised vectors are not ordered, von Neumann pointed at the fact that projectors are  ordered naturally \cite{vN, BvN}: 
\[
P\leq P'\ \Leftrightarrow\ P\circ P' = P
\]
By rescaling density matrices we get a partial ordering that extends these projectors and that can then be used to model the hierarchies mentioned above \cite{bankova2019graded}.  Earlier work used projectors for the same purpose \cite{EsmaSC, BalkirKartsaklisSadrzadeh2015}, and  many years before Widdows \cite{Widdows, widdows2003word} used projectors to encode negation of meanings.  We again use a hand-crafted idealised example for showing how projectors can be used to encode hierarchical relationships:
\beqa 
\rho_{\mbox{\scriptsize\tt lion}} &=& |0\rangle\langle0|\\
\rho_{\mbox{\scriptsize\tt big cat}} &=& |0\rangle\langle0| + |1\rangle\langle1|\\
\rho_{\mbox{\scriptsize\tt mammal}} &=& |0\rangle\langle0| + |1\rangle\langle1| + |2\rangle\langle2|\\
\rho_{\mbox{\scriptsize\tt vertebrate}} &=& |0\rangle\langle0| + |1\rangle\langle1| + |2\rangle\langle2| + |3\rangle\langle3|
\eeqa
Then for example, we could have: 
\[
\rho_{\mbox{\scriptsize\tt tiger}} = |+\rangle\langle+|
\qquad\qquad
\rho_{\mbox{\scriptsize\tt cheeta}} = |-\rangle\langle-|
\]
Moving from projectors to density matrices allows for more gradual transitions. 
\eit    
More recently there has been increased activity in using density matrices for DisCoCat-NLP \cite{CoeckeMeich, MarthaDot, MarthaNeg, GemmaMartha, MarthaNeural}.  Some of these papers address issues that are of foundational physics interest, for example in \cite{CoeckeMeich, MarthaDot, GemmaMartha} the question is asked how to update a density matrix representing a word, given another density matrix, representing an adjective.  Similar questions had been asked earlier in quantum foundations when considering causal inference within the quantum context, mostly be Leifer and Spekkens \cite{Leifer1, leifer2013towards, CoeckeSpekkens2012, PhysRevX.7.031021}.  The recent paper \cite{MarthaNeural} presents three neural models for learning density matrices from a corpus. 

Since we now may want to use density matrices instead of vectors, shall we start all over again?  No, everything stays exactly the same as it was, that is to say, we can still use exactly the same diagrams.  The only difference is that if we now see a state in the diagram then we take it to be a density matrix  instead of a vector: 
\[
\  \longleftrightarrow\  \rho_{\tt hat}
\]
and the application of the grammar map $f_{\cal G}$ proceeds as follows:
\[
f_{\cal G} \circ \left(\rho_{{\tt w}_1}\otimes\ldots\otimes\rho_{{\tt w}_N}\right) \circ f_{\cal G}^\dagger
\]
There is a better manner for representing density matrices which directly matches the diagrammatic representation, for which we refer the reader to \cite{CKbook} Chapter 6.

\paragraph{Complex numbers.}  Also complex numbers have already been used in NLP, for example, in \cite{blacoe2013quantum} where density matrices with complex entries are used as a means to extend the parameter space.  In  \cite{trouillon2016complex, trouillon2017knowledge}  complex numbers are used to factorise knowledge graphs.  So here the motivation for using complex numbers has purely analytical grounds. 
 
\paragraph{Truth values as meanings.}  The internal wiring for transitive verbs that we explained above is indeed very natural.  It also dictates what we can understand by the meaning of a sentence, for example, in the case of a simple sentence with a subject, a transitive verb, and an object: the meaning of a sentence is the meaning of the subject and the object, after these have been updated by the meaning of a verb.  As shown in \cite{CoeckeText, BVgram}, this understanding of what a sentence means is not specific for sentences with a particular grammatical structure, but can be adopted for any sentence.  Recalling a slogan from \cite{CoeckeText}:
\begin{center}
\fbox{\em A sentence is a process that alters the meanings of its words.\em}   
\end{center}
So that's it then?  Have we decided for once and for all what we mean by the meaning of a sentence?  No, there are other possible choices for what we mean by the  meaning of a sentence, which also make a lot of sense, in particular when having some specific tasks in mind. One alternative, already considered in the original DisCoCat paper \cite{CCS}, is to take truth values as sentence meanings, that is, the meaning of a sentence is the degree to which it is truthful. 

The good news is that this case again doesn't require a separate treatment, but is obtained from our previous account on sentence meanings simply by deleting the outputs: 
\[
\input{truth1.tikz}
\]
So we obtain a diagram without either inputs or outputs, which is a number.  If one is worried that this number may be complex, we simply take the square-norm:
\[
\left|\ \input{truth2.tikz}\ \right|^2 
\]
In order to justify this diagram as a degree of truth of the sentence, we first transpose the nouns, by which we mean the following:
\[
\input{truth3.tikz}
\]
Substituting this in the diagram we obtain an inner-product:\footnote{We are being a bit sloppy here about the distinction between the adjoint and the transpose.  If we are only dealing with real-valued matrix entries this doesn't matter, as the adjoint and the transpose then coincide.  Otherwise, Chapter 4 of the book \cite{CKbook} explains the differences very clearly, and how to proceed.}
\beq\label{eq:best-match}
\input{truth4.tikz}\ \ = \ \ \left\langle\ \input{truth5.tikz}\ \right\rangle   
\eeq
So what is this inner-product telling us?  It measures the extent to which the pair ${\tt Alice}\otimes{\tt Bob}$ matches the relationship {\tt hates}.  If this isn't already sufficiently compelling for the fact that in this manner we indeed obtain the degree of truth of the sentence, we again can use a hand-crafted idealised example  to make the case.  Following \cite{CCS, CSC} we can construct a verb like {\tt hates} by superposing all subject-object pairs who satisfy the {\tt hates}-relationship:     
\[
|\psi_{\tt hates}\rangle \ =\ {\tt Alice}\otimes{\tt Bob} + {\tt Wicked~Queen}\otimes{\tt Snow~White}+ {\tt Man.~U.~fan}\otimes{\tt Liverpool} + \ldots
\]
Then we can compute the inner-products: 
\[
\left\langle\, {\tt Wicked~Queen}\otimes{\tt Snow~White} \mid\psi_{\tt hates}\right\rangle\ = \ 1 
\qquad\qquad
\left\langle\, {\tt Romeo}\otimes{\tt Juliette} \mid\psi_{\tt hates}\right\rangle\ = \ 0
\] 
which tell us that {\tt Wicked~Queen hates Snow~White} is very true, while {\tt Romeo hates Juliette} is not at all true.  We can of course also add terms to $|\psi_{\tt hates}\rangle$ which don't stand for full-blown hatred but some degree of 
dislike, for example:
\beqa
\left\langle\, {\tt Bob}\otimes{\tt English\ weather} \mid\psi_{\tt hates}\right\rangle &=& {1\over 3}\\ \ \\ 
\left\langle\, {\tt Bob}\otimes{\tt English\ beer} \mid\psi_{\tt hates}\right\rangle &=& {1\over 2}\\ \ \\
\left\langle\, {\tt Bob}\otimes{\tt English\ chocolate} \mid\psi_{\tt hates}\right\rangle &=& {2\over 3}
\eeqa

\subsection{Quantum-native and quantum advantage}\label{sec:QAdv}  
  
As explained in Section \ref{sec:interaction}, what quantum theory and natural language share at a fundamental level is an interaction structure.   This interaction structure, together with the specification of the spaces where the states live,  determines the entire structure of processes of a theory.  So the fact that  quantum theory and natural language also share the use of vector spaces for describing states---albeit for very different reasons---makes those two theories (essentially) coincide.  Therefore we say that QNLP is `quantum-native'.  What this implies is that natural language  doesn't naturally fit on classical hardware, but instead wants to be on quantum hardware, for the obvious reason that we have adopted a quantum model for language.  This is somewhat similar to simulation of quantum systems having been identified as a problem that would require something like a quantum computer \cite{FeynmanQC}.  But this of course hasn't stopped people from doing simulation of quantum systems on classical hardware, and similarly, we implemented our quantum model of language on classical hardware as well. 

Of course, with the gradually increasing capabilities of quantum hardware, and the growing optimism about practical quantum computing becoming an actuality within the next couple of years, it now is well-motivated to ask the question whether we can think of DisCoCat and related tasks as quantum algorithms, and what the advantages are of doing so.  Obviously, whatever exponential space blow up we have due to the tensor product would immediately vanish.  This point was made in \cite{WillC} by Zeng and Coecke by means of the following table for the particular quantum algorithm that they proposed (D = dimension): 
\begin{center}
\begin{tabular}{c|c|c|c|}
                       &  1 verb, nouns 2K D & 10K verbs, nouns 2K D & 10K verbs, nouns 1M D \\
\hline
 Classical &  $8 \times 10^9$ bits &  $8 \times 10^{13}$ bits &  $8 \times 10^{22}$ bits\\
\hline
 Quantum &   33 qubits     &     47 qubits    &     73 qubits     \\ 
\hline
\end{tabular}    
\end{center} 

They also addressed in \cite{WillC} whether DisCoCat executed on quantum hardware would lead to quantum advantage in terms of speed-up, and with a bit of care DisCoCat would indeed benefit from substantial Grover-like quantum speed-up.  Moreover, meanwhile it has been shown that a number of other standard NLP tasks enjoy quantum advantage \cite{LuongoBellante}. 

As an example we present here a generalisation of the kind of algorithm that was put forward in the first QNLP paper \cite{WillC}, and these tasks are also in line with what we have executed on quantum hardware \cite{QNLPmedium, Nature}. The kind of problem is question-answering, which was considered within the context of (pre-QNLP) DisCoCat in \cite{CDMT, AlexisMSc, de2019functorial}. We will exploit the quantum advantage for computing the closest vector \cite{wiebe2015quantum}.
Concretely, given a vector $|\psi\rangle$, which vector in some set $\{|\psi_i\rangle\}_i$ is closest to $|\psi\rangle$, i.e.~which $i$  maximises $|\langle\psi | \psi_i\rangle|^2$?  

\paragraph{Formulating questions.}  The most common kind of questions are of the form:   
\begin{center}
{\tt Wh*** blablabla?}  
\end{center}
where {\tt Wh***} is either {\tt Who}, {\tt What}, {\tt Which} etc.  One way to think of these  question-words is as a `hole' waiting to be filled-in by the answer.  We will represent such a hole as a `negative box': 
\[
\input{question1.tikz} 
\]
which we did here for the question: {\tt Who hates Bob?}  We can also exchange the roles of subject and object with regard to what is being questioned: 
\[
\input{question2.tikz}  
\]
This question can be thought of as: {\tt Who does Alice hate?}  One can even consider questions which linguistically are more difficult to phrase, but make sense:  
\[
\input{question3.tikz}   
\]
In the first of these `questions' one asks for the verb that best relates the subject and the object, while in the 2nd one of these one asks for the subject-object pair that best matches the verb. Some even more radical variations on the same theme can also be put forward:
\[
\input{question4.tikz}      
\] 
These `questions' correspond to classification tasks: we want to know which noun (typically from some given set of candidate nouns)  best matches the sentence. For example, if this were a newspaper article, would it either be about {\tt sports}, {\tt politics}, {\tt art}, {\tt romance} etc.  Rather than one, we obtain two natural classification tasks, one from the subject's point of view, and one from the object's point of view. If we were to classify the sentence {\tt Alice outperforms Bob} in terms of {\tt winner} or {\tt loser}, the two points of view would yield opposite classifications.   

\paragraph{Answering questions.} In order to answer the questions posed above, one proceeds as follows.  One deletes all remaining outputs, which is equivalent to taking sentence meanings to be truth values, just as we discussed at the end of Section \ref{sec:class}: 
 \[
\input{question5.tikz} 
\]
The optimal answer to this question is the subject that best matches it, that is, results in the most truthful sentence. Hence, we are looking for the subject that maximises:
\[
\left|\ \input{question6.tikz}\ \right|^2
\]
This is an instance of the closest vector problem mentioned above \cite{wiebe2015quantum}, in the following manner:
\[
\mbox{MAX}\left\{ \ 
\left|\left\langle \input{question7.tikz} \right\rangle\right|^2  
\ \begin{tikzpicture}[scale=0.80]
	\begin{pgfonlayer}{nodelayer}
		\node [style=none] (0) at (0, 1.75) {};
		\node [style=none] (1) at (0, -1.75) {};
	\end{pgfonlayer}
	\begin{pgfonlayer}{edgelayer}
		\draw [style=swap] (0.center) to (1.center);
	\end{pgfonlayer}
\end{tikzpicture} \ \
\begin{tikzpicture}[scale=0.80]
	\begin{pgfonlayer}{nodelayer}
		\node [style=none] (0) at (0, -1) {};
		\node [style=none] (1) at (0, -1) {};
		\node [style=none] (2) at (0, 0) {};
		\node [style=langstate, fill=black] (3) at (0, 0.25) {\bW\bf\Large ?\e};
	\end{pgfonlayer}
	\begin{pgfonlayer}{edgelayer}
		\draw [style=swap] (1.center) to (2.center);
	\end{pgfonlayer}
\end{tikzpicture} \in\left\{\input{question10.tikz}, \input{question11.tikz}, \input{question12.tikz}, \ldots  \right\}
\right\}
\]
Hence this question-answering task enjoys quantum speed-up.  The same  would be the case for the classification task 
for which we have:
 \[
\input{question5bis.tikz} 
\]
so where we are looking for the classifier that maximises:  
\[
\left|\ \input{question13.tikz}\ \right|^2
\]
This is an instance of the closest vector problem in the following manner: 
\[
\mbox{MAX}\left\{ \ 
\left|\left\langle \input{question14.tikz} \right\rangle\right|^2  
\ \begin{tikzpicture}[scale=0.80]
	\begin{pgfonlayer}{nodelayer}
		\node [style=none] (0) at (0, 2) {};
		\node [style=none] (1) at (0, -2) {};
	\end{pgfonlayer}
	\begin{pgfonlayer}{edgelayer}
		\draw [style=swap] (0.center) to (1.center);
	\end{pgfonlayer}
\end{tikzpicture} \ \
 \in\left\{\input{question15.tikz}, \input{question16.tikz}, \input{question17.tikz}, \ldots  \right\}
\right\}
\]

It should be clear that the algorithm presented here covers a  wide range of essential NLP tasks.  That said, let us be also blunt about the fact that we really haven't put much effort yet in looking for other NLP tasks that would enjoy different kinds of quantum advantage, possibly even exponential.  This effort, now that we have established proof-of-concept implementations on quantum hardware, will become a major focus for us.  

\section{Grammar+meaning as quantum circuits}\label{sec;QNLPfriendly} 

The diagrams that we have seen thus far represent language and at the same time constitute quantum processes.  However, these quantum processes cannot be directly implemented on existing quantum hardware.  The kind of quantum hardware that we will focus on here are  implementations of the circuit-model for quantum computing.  In the circuit model one may have different sets of basic gates, and here we will assume the basic gates to be phase-gates and CNOT-gates.  So the task at hand is to turn our language diagrams into quantum circuits with phase-gates and CNOT-gates as the only gates.     

For the manner in which the circuits were obtained for the experiments in \cite{QNLPmedium, Nature} we refer to \cite{QPL-QNLP}, as well as for an alternative way of obtaining circuits that stays closer to the original proposal in \cite{WillC}.  While for the experiments done in \cite{QNLPmedium, Nature} both of these approaches are well-suited to the task at hand, they both have one major flaw; namely, the resulting circuits are not compositional.  What we mean by this is that, just like we string sentences together to form text, we also want to be able to composing the corresponding circuits, with the resulting circuit corresponding to the meaning of the resulting text.  Therefore, we won't take the proposals of \cite{QPL-QNLP} as our starting point, but instead, do something entirely new. 

Admittedly, DisCoCat as we presented it above also doesn't allow for composing sentences, as it just produces states.  An alteration of DisCoCat that allows one for composing sentences, named DisCoCirc, was introduced in \cite{CoeckeText}, and was further elaborated upon in \cite{CoeckeMeich}. Very recently, in \cite{BVgram}, a canonical manner to turn grammar into circuits has been put forward, and this result will provide the general template for our discussion here.   

In fact, the analysis in \cite{BVgram} shows that circuits are a more natural way to represent grammar than what we have been presenting above as grammatical structure.  The crux of the argument is that a lot of the complexity of wirings like in (\ref{eq:BIG}) is due to the inability of us humans to speak/hear language beyond 1D strings, poor creatures that we are.\footnote{This argument is similar (although somewhat more subtle) to why diagrammatic reasoning is also much more appealing and easier  to us than corresponding symbolic reasoning, for which the case is made in Section 3.2.4 of \cite{CKbook}.}   Hence, in particular, there is no reason why a machine like a computer should deal internally with language in the same unnecessary cumbersome way as us humans deal with it. We claim that a machine should deal with language as circuits instead.  If the reader buys into this argument, then  quantum computational circuits are particularly friendly towards natural language, in circuit-shape of course.   That said, in Section \ref{sec:variationalCirc} we provide a much an even more powerful argument in favour of NISQ being QNLP-friendly, building further on this language as circuits paradigm. 


\subsection{Compositional language circuits}\label{sec:complangcirc} 

\paragraph{Z- and X-spiders.}  Below we will make use of two special kinds of spiders, namely the ones that make up the  ZX-calculus \cite{CD1, CD2}. These respectively arise from the Z- and the X-bases:
\beqa
\input{spidercomp.tikz}\ &=& \ |0\ldots 0\rangle\langle 0\ldots 0| + |1\ldots 1\rangle\langle 1\ldots 1|\\
\input{spidercomp2.tikz}\ &=& \ |+\ldots +\rangle\langle +\ldots +| + |-\ldots -\rangle\langle -\ldots -|  
\eeqa
Two special spiders that we will use frequently are:  
\[
\begin{tikzpicture}[scale=0.80]
	\begin{pgfonlayer}{nodelayer}
		\node [style=none] (0) at (0, -0.5) {};
		\node [style=gray dot] (1) at (0, 0.75) {};
		\node [style=none] (2) at (0, 0.75) {};
	\end{pgfonlayer}
	\begin{pgfonlayer}{edgelayer}
		\draw [style=swap, in=-90, out=90, looseness=0.75] (0.center) to (1);
	\end{pgfonlayer}
\end{tikzpicture}\ \ = \ \sqrt{2}\,| 0\rangle
\qquad\qquad\qquad\qquad
\begin{tikzpicture}[scale=0.80]
	\begin{pgfonlayer}{nodelayer}
		\node [style=none] (0) at (0, 0.75) {};
		\node [style=gray dot] (1) at (0, -0.5) {};
		\node [style=none] (2) at (0, -0.5) {};
	\end{pgfonlayer}
	\begin{pgfonlayer}{edgelayer}
		\draw [style=swap, in=90, out=-90, looseness=0.75] (0.center) to (1);
	\end{pgfonlayer}
\end{tikzpicture}\ \ = \  \sqrt{2}\,\langle 0|
\]
Most of the time we will ignore the presence of  scalars like $\sqrt{2}$, since as explained in \cite{CKbook}, it is easy to recover them at the end of a diagrammatic calculation.  Our main use in this section of these differently coloured spider is that if we plug them together as follows:  
\[  
\input{cnot.tikz} 
\]
we obtain the CNOT-gate, which is easily checked. If the horizontal wire confuses the reader, then they can think of it as either of the following two alternatives, which turn out to be equal anyway: 
\[  
\input{cnotbis.tikz} 
\]
What is key here is that logically speaking, the CNOT-gate should not be conceived as a monolithic whole, but as something that is made up of two smaller entities, namely two differently coloured spiders.  As we shall see below, this ZX-formulation of the CNOT-gate is crucial for moving from the quantum representations of natural language to a circuit-form that is acceptable by existing hardware.  

\par\bigskip  

Picking up where we left things in Section \ref{sec:class},  we are in fact already quite close to a circuit representation.  The trick is to now pull some more spiders out of our sleeve:
\beq\label{eq:tv5} 
\input{tv5.tikz}   
\eeq
Let's make it a bit clearer what exactly happened here:
\[
\input{pullout.tikz}
\]
In (\ref{eq:tv5}) there now are a number of prepared states, {\tt Alice}, {\tt *hates*} and {\tt Bob}, two CNOT-gates, and two grey deletes, which practically boils down to a post-selected measurement:
\[  
\input{tv6.tikz}   
\]

\paragraph{Composing sentences.} We now show how sentences like this one, once they are in this circuit form, can be composed in order to account for larger text.  Besides the previous sentence, assume that we also have:
\beq\label{eq:tv7}
\input{tv7.tikz}   
\eeq
This is how we put these two sentences together:    
\beq\label{eq:tv8}
\input{tv8.tikz}   
\eeq
The key point here is that, while we start with {\tt Alice} and {\tt Bob} as separated states, after we learned that {\tt Alice *hates* Bob}, they have become entangled, and in particular, the meanings of {\tt Alice} and {\tt Bob} have been altered: while we started off with meanings of {\tt Alice} and {\tt Bob} that essentially contained no knowledge about them whatsoever, after the 1st sentence we learned something about both of them, namely the nature of their relationship, so their meanings indeed underwent a change. It are these altered meanings, and {\tt Bob}'s in particular, to which the circuit representing the 2nd sentence is then applied. 

Let's go for a more sophisticated example like {\tt Alice hates Bob who likes beer}.  This sentence includes two clearly identifiable parts: the sentence  {\tt Alice hates Bob} and the noun-phrase {\tt Bob who likes beer}. Let's see if we can get the latter also in circuit form.  Substituting the internal wiring (\ref{eq:verb}) in the noun-phrase wiring (\ref{eq:relpronQNLP}) we obtain:
\beq\label{eq:relproncircuit}
\input{relproncircuit.tikz} 
\eeq
We can  reshape this into:     
\[  
\input{tv9a.tikz}
\]
and again we pull some spiders out of our sleeve: 
\[  
\input{tv9b.tikz}
\]
This resulting circuit is not entirely satisfactory as a later sentence may be more specific about which kinds of beer Bob likes, so we want to be able to let the meaning of beer evolve further.  Hence, we need to be able to compose the circuit-wire representing beer with additional sentence-circuits.  Therefore,  we simply get rid of the white delete and hence obtain (\ref{eq:tv7}).
This is in accord with our  more sophisticated sentence exactly conveys the same information as the two sentences in the circuit (\ref{eq:tv8}).\footnote{The `hack' of  `un-deleting' {\tt beer} boils down to fixing a shortcoming in language-diagrams like (\ref{eq:relproncircuit}), and this shortcoming vanishes when moving to the circuit form \cite{BVgram}.  As already mentioned earlier, we indeed like to think of circuits as a more foundational presentation of grammar.} 

\paragraph{Generalisation.}  The manner in which we turned grammar diagrams into circuits for our particular example, generalises to a recipe that works for all of English, and pretty much any existing language in the world.  We won't go into any more details here, and refer the interested reader to \cite{BVgram}.  All together we will either obtain 1-ary (e.g.~adjective), 2-ary (e.g.~transitive verb), and 3-ary  (e.g.~ditransitive verb) gates acting on noun-wires:   
\[  
\input{tv18.tikz}  
\]

\paragraph{Parallelisation vs.~sequentialisation.} The circuit (\ref{eq:tv7}) requires 4 qubits and has two CNOT-gates in parallel. We can also derive an alternative circuit that reduces the number of qubits, but increases the depth of the CNOT-gates.  Which circuit that one prefers ultimately depends on the hardware that one uses, whether it has plenty of qubits available, or, whether it allows for implementing CNOT-gates sequentially with high fidelity. For example, among the currently available hardware, superconducting hardware (e.g.~IBM, Google, Rigetti) has plenty of qubits, but performs badly for circuit depths that are too large, while ion trap hardware (IonQ, Honeywell, Universal Quantum, Oxford Quantum Circuits) comes with less qubits, but does better for greater circuit depth.  

Using the Choi-Jamiolkowski correspondence again: 
\[
\input{CJ2.tikz}
\]
we obtain:
\beq\label{eq:tv12}    
\input{tv12.tikz}   
\eeq
So we eliminated one qubit at the cost of CNOT-gates requiring sequential application.


\subsection{Modelling verbs and nouns} 

\paragraph{ZX-phases.}  In order to achieve universality with respect to qubit quantum computing, in full-blown ZX-calculus \cite{CD1, CD2}  spiders also cary `decorations', a.k.a.~`phases': 
\beqa
\input{spidercompalph.tikz}\ &=& \ |0\ldots 0\rangle\langle 0\ldots 0| + e^{i\alpha}|1\ldots 1\rangle\langle 1\ldots 1|\\
\input{spidercomp2alph.tikz}\ &=& \ |+\ldots +\rangle\langle +\ldots +| + e^{i\alpha} |-\ldots -\rangle\langle -\ldots -|  
\eeqa
These spiders just compose as usual, provided one adds  the phases:
\[ 
\input{spiderphase.tikz}\ \ =\ \ \input{spidercompphase.tikz}
\] 
Phase gates are special cases of the above spiders:
\[
\input{spidercompgate.tikz}\ = \ |0\rangle\langle 0| + e^{i\alpha}|1\rangle\langle 1|\qquad\qquad
\input{spidercomp2gate.tikz}\ = \ |+\rangle\langle +| + e^{i\alpha} |-\rangle\langle -|  
\]
Hence, in ZX-calculus we now have the ability to write down the CNOT-gate (as we saw above), and via Euler decomposition, also any arbitrary one-qubit unitary gate:
\[
\input{unitary.tikz}
\]
Hence, we obtain a universal language. It is indeed well known that CNOT-gates and arbitrary one-qubit unitary gates together generate any arbitrary unitary circuit.\footnote{This argument can be easily extended to showing that ZX-calculus is universal for all linear maps \cite{ContPhys, CD2}.} 

\par\bigskip 

Above  verbs like {\tt*hates*} and {\tt*likes*} were still depicted as `black boxes' rather than having a circuit representation.  The same was also the case for the nouns {\tt Alice}, {\tt Bob} and {\tt beer}.  We will now also turn these into circuits.  Here we will take the form (\ref{eq:tv12}) as our starting point, but everything that we say here can also be done in terms of the parallelised form (\ref{eq:tv5}).  

Having in mind that we are still in the NISQ-era, rather than going for the most general representation, one may wish simplify things a little bit, likely depending on the task at hand.  Here we present some options for how one could model these verbs and nouns.\footnote{Earlier experiments that were concerned with simplifying word representation in DisCoCat, although not with the aim to produce circuits, can be found in \cite{Kartsaklis13reasoningabout}.}  

We start with an almost trivial representation of verbs, which nonetheless is useful for certain applications.  We represent all verbs by Bell-states:
\[
\input{CJ3.tikz}
\]
Through the Choi-Jamiolkowski correspondence this becomes: 
\[
\input{CJ4.tikz}
\]
The circuits for {\tt Alice hates Bob} as well as the one for {\tt Alice loves Bob} then both become:
\[
\input{tv13.tikz}
\]
So sentences with opposite meanings end up being assigned the same circuit.  Does this have any use?  Yes, if what we are concerned with is whether there is any kind of relationship between {\tt Alice} and {\tt Bob}, then this diagram tells us everything we need to know.  Of course, while not entirely useless, this representation for verbs would only have limited applicability.

Instead of by an identity, let us represent verbs by an arbitrary unitary gate $U$, then using the Euler decomposition---mentioned earlier in this section:  
\[
\input{tv14.tikz}
\]
We now have a large space of possible verb-meanings that allow one to distinguish {\tt*hates*} and {\tt*likes*}.  Still, while each verb may have different values for $\alpha$, $\beta$ and $\gamma$, all verbs are equally connecting the subject and the object, or equivalently, in terms of states, maximally entangling.  One may argue that this should not be the case for example for {\tt knows} vs.~{\tt marries}.   For a general representation for the verb, we can rely on singular value decomposition:
\[
\input{tv15.tikz}
\]
where {\bf p} represents the diagonal of the diagonal matrix:
\[
\input{tv16.tikz}\ \ = \ \ {\bf p}_0 |0\rangle\langle 0| + {\bf p}_1 |1\rangle\langle 1|
\]
where again we pulled some spiders out of our sleeve in order to obtain a circuit form.

In order to complete the circuit picture, we can now also replace all noun states by gates:
\[
\input{tv17.tikz}
\]
As a result we obtain a circuit only consisting of CNOT-gates and phase gates: 
\[
\mbox{CNOT}\ = \ \ \input{cnot.tikz}
\qquad\qquad
\mbox{Z-phase}\ = \ \ \input{spidercompgate.tikz}
\qquad\qquad
\mbox{X-phase}\ = \ \ \input{spidercomp2gate.tikz}
\]
Moreover, using Hadamard gates all phases can become Z-phases:  
\[
\input{spidercomp2gate.tikz}\  = \ \ \input{spidercompgateH.tikz}
\]
 
\paragraph{Complete ZX-rules.}  Above we made use of a small fragment of the ZX-calculus.  Even when it was introduced in \cite{CD1, CD2} the ZX-calculus had a number of additional rules, which even today are still the ones that are being used in practice, e.g.:  
\[
\input{ZXrules.tikz}
\]
Recently it was shown that ZX-calculus when some extra rules are added---in the most recent result, a single rule in fact---can reproduce any equation that one can establish using linear algebra \cite{hadzihasanovic2018two, vilmart2018near}.  In logic-terms this is referred to as the ZX-calculus being `complete' for qubit quantum computing.  This fact even more justifies our representation of natural language, not only in terms of a quantum model, but in terms of ZX-calculus in particular---see also Section  \ref{sec;scalemean} below where we consider higher dimensional meaning spaces.

\paragraph{Circuit optimisation using ZX-calculus.}
Also directly relevant here is the fact that ZX-calculus has meanwhile became part of state-of-the-art for automated quantum circuit optimisation \cite{duncan2019graph, kissinger2019pyzx, kissinger2019reducing, de2020fast}.  This allows our circuits representing language to be further simplified, and hence allows one to increase the size of the tasks that we can do on existing hardware.  The size-reduction itself is build-in in Cambridge Quantum Computing's t$|$ket$\rangle$ compiler \cite{sivarajah2020t}, which in our experiments played the role of the interface between our software and the hardware. In order to make t$|$ket$\rangle$ understand our language diagrams, we relied on the DisCoPy toolbox \cite{DisCoPy}.   

\subsection{Varying the basis}\label{sec:varybasis}  

The cautious reader may have noticed that in pictures like this ones:
\[
\input{tv5basis.tikz}   
\]
we used a fixed basis to specify the spiders.  Spiders are  highly basis-dependent, to the extend that they specify an orthonormal basis \cite{CPV}.  In order to see this, it suffices to note that the copying-spider copies basis vectors, and hence by the no-cloning theorem, cannot copy any other vectors.  Hence, it specifies an orthonormal basis as the vectors that it copies.  

There are two possible narratives to address this. One, which essentially amounts to discarding the issue, goes as follows:
\bit
\item A good portion of NLP that relies on `hand-crafted' meaning spaces follows Firth's dictum \cite{Firth} ``you shall know a word by the company it keeps", and many of these approaches, most notably the earlier mention bag-of-words model \cite{harris1954distributional}, are highly basis-dependent by construction.  For example, following \cite{Schuetze}, one freely spans an orthonormal basis by chosen `context-words' $\{{\tt y}_i\}_i$, and meanings of other words {\tt x} are then crafted simply by counting how  many times {\tt x} appears close to each of the ${\tt y}_i$.  The  relative frequencies of these co-occurences then become the coordinates relative to that orthonormal basis.    
\eit
More recent approaches to NLP have moved away from this principle.  Therefore, rather than ignoring the issue that there is a preferred orthonormal basis, is to take the basis to be a variable.  This then also provides additional parameter flexibility.  Let us  indicate the variable basis  using a different colour, here black:
\[
\input{tv5basis2.tikz}     
\]
From the definition of spiders, it easily follows that when varying the basis they transform  as follows:
\[
\input{spidertransfpre.tikz}     
\] 
for some one-qubit unitary gate $U$, so we can parametrise the varying basis using the Euler decomposition in terms of phases: 
\[
\input{spidertransf.tikz}     
\] 
Hence we obtain the following general form:
\[
\input{tv5basis3.tikz}       
\] 
where the $\gamma$-phase and -$\gamma$-phase cancel out, and {\tt *hates*} absorbs some of the phases, in that process becoming {\tt **hates**}.

\subsection{Higher dimensional meaning spaces}\label{sec;scalemean} 

For the sake of simplicity of the argument we assumed noun-spaces to be $\mathbb{C}^2$. Evidently, in practice one would go well beyond small dimensions like this.  In order to achieve that one could just string qubits together until one reaches the desired dimension, and treat the resulting space just as a Hilbert space without any further structure.  However, doing so one misses out on a great opportunity, namely, the opportunity  to craft a meaning model that truly embraces the quantum nature of the hardware.  Indeed, we already saw that the quantum model  allows one to combine meaning and linguistic structure like grammar, but also within the meaning spaces themselves we can make quantum structure play a significant role.  

We will illustrate this by means of an example initially introduced in \cite{ConcSpacI}.  Let's think a bit about the meaning of {\tt banana}.  There are a few features that can be attributed to  {\tt banana}, or to fruit in general. One is its colour, one is its taste, another one is texture, and so on.   But can we simply say that a banana is yellow?  Not really, since when you buy them they may be green, and if you don't eat them in time they turn black.  So the better way to represent the meaning of {\tt banana} would be as follows:
\[
|\psi_{\tt banana}\rangle\  = \ 
|{\tt green}\rangle|{\tt bitter}\rangle|{\tt hard}\rangle
+
|{\tt yellow}\rangle|{\tt sweet}\rangle|{\tt soft}\rangle 
+
|{\tt black}\rangle|{\tt yucky}\rangle|{\tt slimy}\rangle
\]
What we are exploiting here is entanglement in order to encode  meaning-correlations. Again, classically such an approach would lead to an exponential explosion of the dimension, while when doing things quantumly, this really seems to be the natural way forward. 

The circuits representing grammar stay the same when we scale up dimension, except for having multiple copies of the same circuit:
\[
\input{tv5mult.tikz}          
\] 
In the light of the discussion in the previous paragraph, we can now think of these different layers as each being restricted to a particular feature like colour, taste, texture, etc.

\paragraph{ZX-calculus for qudits.} The use of ZX-calculus above doesn't make any essential use of qubit-specific rules, so our translation from language to quantum circuits can equally well be done for qudits.  For the specific case of qutrits ZX-calculus is already well-developed  \cite{wang2014qutrit, wang2018qutrit}, and higher dimensional variants are currently also being developed, most notably by Quanlong (Harny) Wang.   

\section{Learning meanings quantumly}\label{sec:variationalCirc}  

A typical first lesson on quantum computing goes as follows: 
``Quantum computers are amazing since thanks to superposition, they can apply a function at once to all possible arguments.  However, unfortunately, we don't have direct access to the result of doing so.'' 
That is, we can't get the data sitting inside a quantum computer out of it like we are used to with classical computers.  What is not part of the lecture, but what is the case for the currently available quantum computers, is that one also cannot get one's data into the quantum computer in the way that we are used to with classical computers.  The tools that would enable one to do so, e.g.~{\color{red} QRAM} \cite{giovannetti2008quantum}---on which \cite{WillC} relied for its argument---haven't  (yet) materialised. Hence, one needs to come up with alternative ways of feeding data into to the quantum computer.

\subsection{Variational quantum circuits}\label{sec:varcirc}

The current leading  paradigm for (indirectly) encoding data on a quantum computer are variational circuits \cite{ma2019variational, benedetti2019parameterized, schuld2020circuit}.  In fact, most people conceive variational circuits as a family of quantum machine learning algorithms which enjoy a number of particular advantages.  

The name variational quantum circuits refers to the fact that one is dealing with quantum circuits in which there are a number of variables, just like the circuits we have been building in the previous sections for our language diagrams, like this one:  
\[
\input{tv14bis.tikz}         
\] 
We now explain how we encode data by means of variational quantum circuits.  

Assume as given some training data.  This could be a training set of sentences, or texts, of which the overall meaning is assumed to be known.  Let's denote the training data by $\{{\tt w}_1^i \ldots {\tt w}_{N_1}^i\}_i$, and the corresponding meanings by $\{|\psi^i\rangle\}_i$:
\[
\input{data.tikz}   
\]
A special case is the one where all the $|\psi^i\rangle$'s are truth values, in the sense discussed in Section \ref{sec:class}, and we could then restrict to all data in the training set being truthful:  
\[
\input{data2.tikz} 
\]
This last case can be efficiently implemented, and hence supports quantum speed-up.\footnote{For more general quantum machine learning, potential quantum speed-ups well beyond Grover-like speed-ups have already been identified, for example in \cite{hastings2020classical, liu2020rigorous, sweke2020quantum}.  As already mentioned before, we also expect post-Grover speed-ups to be achievable for QNLP.}  For the sake of generality we outline a procedure here that applies to general meanings.

First we assign to each sentence ${\tt w}_1^i \ldots {\tt w}_{N_1}^i$ of the training data an associated circuit:
\[
\left\{   
{\tt w}_1^i \ldots {\tt w}_{N_1}^i\  \mapsto \ 
\input{abs2i.tikz}
\ \right\}_i 
\]
just as we did in the previous sections.  Next we assign some initial settings to the variables $\alpha^i, \beta^i, \ldots$ in the circuit. 
We can now start `training' the variables by means of some classical optimisation algorithm aiming to match the training data as good as possible. More specifically, we repeat the following steps:
\ben
\item Measure the circuits following some procedure, resulting in $\{|\tilde\psi^i\rangle\}_i$.  
\item Re-adjust the variable settings accordingly, modifying each by $\{\delta\alpha^i, \delta\beta^i, \ldots\}_i$. 
\een 
\[
\input{optimise.tikz}
\]
In the case of general meanings this would require a tomographic procedure, while for truth-valued it suffices to maximise a positive measurement outcome for $|+\ldots +\rangle$:  
\[
{\input{abs2ibis.tikz}}
\]
We continue to repeat the steps indicated above until an optimal setting for all variables is reached with respect to the training data.  After doing so we obtain circuit variables settings for all the word meanings that occur in the training set of sentences.  This is a crucial point: meanings of words are not stored as data in the quantum computer, but they correspond to classically controlled  circuit variables, and we learn these with the help of an optimisation algorithm from the training data.  So in fact, the word meanings come in the form of classical  control data $\alpha^i, \beta^i, \ldots$, not as quantum data.  

Once the data is encoded in the quantum computer in this manner, we can unleash our quantum algorithms, for example those discussed in section \ref{sec:QAdv}. For the experiments that we have already did \cite{QNLPmedium, Nature}, we did question-answering. 

\subsection{NISQ is QNLP-friendly} 

 Now here is the real upshot about the variational quantum circuits paradigm from our particular perspective.  The starting point for variational quantum circuits are quantum circuits in which there are a number of variables.  So the first question is: How do we pick these quantum circuits?  For most areas that employ variational quantum circuits there is no canonical choice, and what guides the decision is mostly how well a particular choice of quantum circuits performs.  Therefore, they are referred to by means of the term `ansatz', acknowledging the fact that the chosen circuit form is rather empirically than conceptually motivated.\footnote{For other uses of variational quantum circuits, certain aspects of the ansatz may reflect things like periodicity, bond dimension, infinity etc., or be characteristic for a certain class of problems, just like it is the case for traditional machine learning where, for example, for images one typically uses convolutional neural networks.}   On the other hand, for us, these circuits representclass o the structure of natural language on-the-nose! Contrast this with our discussion of Section \ref{sec:class} of DisCoCat when executed on a classical hardware, where accounting for the same structure was exponentially expensive.   

So to recap, the circuits that come as `part of the package' of variational quantum circuits, are linguistic structure,
in contrast to the apparently  exponentially expensive classical encoding of grammar,  and as compared to most other NISQ-practitioners of the paradigm, an ansatz is replaced by a fully meaningful structure.  Hence, we truly get a free lunch.
  
Now, looking way back in time, some 50 years ago, it has been argued that grammar is somehow hard-wired within our brains, by the likes of Chomsky \cite{Chomsky65} and Montague \cite{montague1970universal}, and even many thousands of years before that by Bacon \cite{Bacon}. We now learned that grammar is indeed hard-wired somewhere, namely in a NISQ quantum computer!


\subsection{Learning parts from the whole}\label{sec:topdown} 

In this last section we wish to emphasise a very important feature of what we just described.  Way back in Section \ref{sec:combiningmeaninggrammar}  we explained how DisCoCat arose from the desire to produce the meaning of a whole, given the meanings of parts.  For example, given the meanings of words it provides an algorithm that produces the meaning of a sentence.  Let's call this  `bottom-up'.    

On the other hand, in Section \ref{sec:varcirc} we did the exact opposite, say `top-down':  given the meanings of the whole, we derived the meanings of the parts.   Our valuable toilet roll specified what we want some sentences to mean, and from this, through an optimisation procedure, we derived  the `best fit'  for the meanings of the words the sentence was made up from.  

This in itself constitutes a  novel approach to grammar-informed NLP.  Let's first recall  Wittgenstein's conception of meaning-is-context, and Firth's derived dictum \cite{Firth} ``you shall know a word by the company it keeps", a prominent  paradigm in the early days of DisCoCat---see Section \ref{sec:varybasis}. For all previous approaches to grammar-informed NLP, one entirely ignored `how' words interacted with each other.  We belief that we are taking a major step forward on this issue by truly accounting for how words interact when establishing their meanings.  Within the same vein, modern machine learning has made an incredible amount of progress.  It did so to a great extend by boosting computational power while treating the actual learning core as some kind of black box.  Instead, we prefer to retain some insight on where in the system the meanings are, evolve, and interact.  Let's call that a meaning-aware system.  We believe that in the long term, this approach will give us a great advantage on the black box approach. Evidence for our stance is provided by the history of science, where all paradigmatic revolutions went hand-in-hand with a new fundamental structural understanding. 

 
\section{Digest}

In this paper we outlined conceptual and mathematical foundations for quantum natural language processing. Admittedly,  we didn't provide full generality on many issues, for example, we only presented how to turn language diagrams into quantum circuits for very specific cases,  and by no means provided anything like an exhaustive account on quantum speed-up for QNLP. That said, the number one message of this paper is that QNLP is not just another quantum counterpart to some classical method.  Instead, already at a foundational level, it provides a unique opportunity for quantum computing, closely related to the opportunity which simulation of quantum systems provides.    We captured this in terms of a number of buzz-words: 
\bit
\item {\bf QNLP being quantum-native}  points at the fact that what canonically combines meaning and linguistic structure into one whole is a quantum model, and hence any task involving that model more has quantum hardware as its natural habitat. 
\item {\bf NISQ being QNLP-friendly}   points at the fact that in the licht of the variational quantum circuits paradigm, when used to encode classical data on quantum hardware, accounting for linguistic structure comes as a free lunch while on classical hardware it looks to be exponentially expensive.
\item {\bf QNLP being meaning-aware} points at the fact that in contrast to much of modern machine learning where  everything takes place inside a black box,  the flows of meanings in QNLP are clearly exposed and understood, and we aim to further develop this perspective, not only exposing the flows of meanings due to the linguistic structure, but also those that happen within the meanings themselves.
\eit 
We also showed that QNLP is based on two foundational observations:
\ben
\item  Due to their common structure, most notably in terms of diagrams, linguistic diagrams can be on-the-nose interpreted in terms of quantum processes: 
\[ 
\input{tv2wg.tikz} 
\ \ \ \mapsto\ \ \ 
\Bigl(
\left\langle Bell\right| \otimes \mathbb{I}\otimes \left\langle Bell\right|
\Bigr)
\circ
\Bigl(
|\psi_{{\tt Alice}}\rangle\otimes |\psi_{\tt hates}\rangle  \otimes|\psi_{{\tt Bob}}\rangle
\Bigr)
\]
\item Using ZX-calculus, these dual linguistic/quantum diagrams can then be brought into circuit form, matching the existing quantum hardware:
\[
\input{tv2wg.tikz} 
\ \ \ \mapsto\ \ \ 
\input{tv5bis.tikz}
\]
\een
Each of these two steps makes use of diagrammatic reasoning in a fundamental way.  

\section{Acknowledgements}   

Alexis thanks Simon Harrison for his generous support for the Wolfson Harrison UK Research Council Quantum Foundation Scholarship that he enjoys.  Konstantinos is grateful to the Royal Commission for the Exhibition of 1851 for financial support under a Postdoctoral Research Fellowship.  We  thank Antonin Delpeuch for his (friendly) mocking of the DisCoCat program for ``[...] only waiting for the right
hardware (such as quantum devices) to be unleashed" in \cite{delpeuch2019autonomization}, which thanks to Ilyas Khan and CQC's initial encouragements is now  happening.  We thank Marcello Benedetti, Ross Duncan and Mattia Fiorentini for  bringing us up to date with the state-of-the-art on QML and available quantum hardware during the early days of this endeavour.  We thank Marcello Benedetti, Ross Duncan, Dimitri Kartsaklis, Quanlong Wang and Richie Yeung for valuable feedback. 

\bibliographystyle{plain}
\bibliography{main}

\end{document}